\documentclass[numreferences]{kluwer}

\usepackage{times}
\usepackage{graphicx}

\usepackage{makeidx}
\def\indexname{Index}
\makeindex

\newcommand{\sqrtsNN}{\mbox{$\sqrt{s_{_{\mathrm{NN}}}}$}}
\newcommand{\pt}{\mbox{$p_\perp$}}

\newcommand{\gevc}{\mbox{${\mathrm{GeV/}}c$}}
\newcommand{\mevc}{\mbox{${\mathrm{MeV/}}c$}}
\newcommand{\jpsi}{\mbox{$J/\Psi$}}
\newcommand{\ep}{\mbox{$\mathrm{e}^+\mathrm{e}^-$}}

\begin{document}
\begin{article}

\begin{opening}
\title{Heavy Ion Experiments at RHIC: The First Year
    \footnote{The original title of the manuscript is: \textit{Experimentalists are from Mars, Theorists are from Venus}}}
\subtitle{\rm Writeup of Lectures given at the NATO Summer School on\\
    \textit{QCD Perspectives on Hot and Dense Matter} in Cargese, France }
\author{J.L. \surname{Nagle}\email{nagle@nevis.columbia.edu}}
\institute{Columbia University, New York, NY 10027, USA}
\author{T. \surname{Ullrich}\email{ullrich@bnl.gov}}
\institute{Brookhaven National Laboratory, Upton, NY 11973, USA}

\runningauthor{J.L. Nagle and T. Ullrich}
\runningtitle{Experimentalists and Theorists}

\begin{ao}\\
KLUWER~ACADEMIC~PUBLISHERS PrePress Department,\\
P.O.~Box~17, 3300~AA~~Dordrecht, The~Netherlands\\
e-mail: TEXHELP@WKAP.NL\\
Fax: +31 78 6392500
\end{ao}
\date{\filedate}

\begin{abstract}
    We present a written version of four lectures given at the NATO
    Advanced Study Institute on ``QCD Perspectives on Hot and Dense
    Matter'' in Cargese, Corsica during August, 2001.  Over the last
    year the first exciting results from the Relativistic Heavy Ion
    Collider (RHIC) and the four experiments BRAHMS, PHENIX, PHOBOS,
    and STAR have been presented.  In these lectures we review the
    state of RHIC and the experiments and the most exciting current
    results from Run I which took place in 2000.  
    A complete review is not possible yet with many key
    results still preliminary or to be measured in Run II, which is
    currently underway, and thus
    the emphasis will be on the approach experimentalists have taken
    to address the fundamental physics issues of the field.  We have
    not attempted to update the RHIC results for this proceedings, but
    rather present it as a snapshot of what was discussed in the
    workshop.  The field is developing very quickly, and benefits
    greatly from contact and discussions between the different
    approaches of experimentalists and theorists.
\end{abstract}

\end{opening}

{\footnotesize\tableofcontents}


\section{Introduction}

A good way to proceed in understanding the construction, operation,
and physics output from current high-energy heavy ion experiments is
to start with the original physics motivations for the experimental
program.  The Relativistic Heavy Ion Collider project was started in
the 1980's with a list of experimental observables for characterizing
hot and dense quark and gluonic matter and the expected restoration of
approximate chiral symmetry and screening of the long range confining
potential of QCD.  We give a brief and certainly not complete review
of these physics signals.  Then we discuss some of the cost, schedule
and technology constraints that impacted the design and construction
of the RHIC experiments.  Finally we present a sample of the first
results from the RHIC experimental program from data taken during Run
I in the summer of 2000.


\section{Physics Motivations}

There are three main categories of observables that were originally
proposed to study the matter produced in relativistic heavy ion
collisions: (1) Thermodynamic properties of the
system and indications of a possible first or second order phase
transition between hadronic matter and quark-gluon plasma, (2)
Signatures for the restoration of approximate chiral symmetry
transition, and (3) Signatures for deconfinement and the screening of
the long range confining potential between color charges.  In the
following list, there will be specific channels that need to be
observed that essentially specify the types of detectors and
experiments necessary.

\subsection{Thermodynamic Properties}

One of the most important question in the physics of heavy-ion
collisions is thermalization.  We want to describe the system in terms
of a few thermodynamic properties, otherwise it is not possible to
discuss an equation-of-state and a true order to any associated phase
transitions.  The use of thermodynamic concepts to multi-particle
production has a long history. One of the first to apply them in
elementary collisions was Hagedorn in the early 1960's
\cite{hagedorn}. The concept of a \textit{temperature} applies
strictly speaking only to systems in at least local thermal
equilibrium. Thermalization is normally only thought to occur in the
transverse degrees of freedom as reflected in the Lorentz invariant
distributions of particles. The measured hadron spectra contain two
pieces of information: \textit{(i)} their normalization,
\textit{i.e.}~their yields ratios, provide the chemical composition of
the fireball at the chemical freeze-out point where the hadron
abundances freeze-out and \textit{(ii)} their transverse momentum
spectra which provides information about thermalization of the
momentum distributions and collective flow. The latter is caused by
thermodynamic pressure and reflects the integrated equation of state
of the fireball matter.  It is obvious that the observed particle
spectra do not reflect earlier conditions, \textit{i.e.} the hot and
dense deconfined phase, where chemical and thermal equilibrium may
have been established, since re-scattering erases most traces from the
dense phase. Nevertheless, those which are accumulative during the
expansion, such as flow, remain.  Only direct photons, either real
photons or virtual photons that split into lepton pairs, escape the
system without re-scattering.  Thus these electromagnetic probes yield
information on the earliest thermodynamic state which may be dominated
by intense quark-quark scattering.

The assumption of a locally thermalized source in chemical equilibrium
can be tested by using statistical thermal models to describe the
ratios of various emitted particles. This yields a baryon chemical
potential $\mu_B$, a strangeness saturation factor $\gamma_s$, and the
temperature $T_{ch}$ at chemical freeze-out.

So far these models are very successful in describing particle ratios
at SPS \cite{pbm-sps} and now also at RHIC \cite{pbm-rhic}.  At RHIC
the derived chemical freeze-out temperature are found to be around 175
MeV (165 MeV at SPS) and a baryon chemical potential of around 45 MeV
(270 MeV at SPS). It should be stressed that these models assume
thermal equilibration but their success together with the large
collective flow (radial and elliptic) measured at RHIC is a strong
hint that this picture indeed applies.

\subsection{Chiral Symmetry Restoration}
The phase transition to a quark-gluon plasma is expected to be
associated with a strong change in the chiral condensate, often
referred to as the restoration of approximate chiral symmetry in
relativistic heavy ion collisions is discussed here.  Note that
although many in the field refer to the restoration of chiral
symmetry, the system always breaks chiral symmetry at a small scale
due to the non-zero neutral current masses of the up and down quarks.

There are multiple signatures of this transition, including disoriented
chiral condensates (DCC), strangeness enhancement, and many others.
However, the most promising signature is the in medium modification to
the mass and width of the low mass vector mesons.  Nature has provided
an excellent set of probes in the various low mass vector meson states
($\rho$, $\omega$, $\phi$) whose mass poles and spectra are
dynamically determined via the collisions of hadrons or partons.  If
the hot and dense state produced in heavy ion collisions is composed
of nearly massless partons, the $\rho$(770) meson mass distribution is
expected to broaden significantly and shift to lower values of
invariant mass.  The $\rho$ meson has a lifetime that is $\tau \approx
1~\mathrm{fm}/c$ and the plasma state created in RHIC collisions has a lifetime
of order 10 fm/$c$.  Thus, the $\rho$ meson is created and decays
many times during the entire time evolution of the collision.

The $\rho$ has a dominant decay (nearly 100\%) into two pions.
However, in this decay mode if the pions suffer re-scattering with
other hadrons after the decay, one cannot experimentally reconstruct
the $\rho$ meson and information is lost.  Given the dense, either
partonic or hadronic environment, the probability for pion
re-scattering is very large, unless the $\rho$ is created and decays
at the latest stages of the time evolution.  This time period is often
referred to as thermal freeze-out, when elastic collisions cease.
Thus a measurement of the $\rho$ as reconstructed via its pion decay
channel gives interesting information on the final hadronic stage, but
not on the dense phase where chiral symmetry may be restored.

There is an additional decay channel into electron pairs and muon pairs, 
though with small branching ratios of $4.5
\cdot 10^{-5}$ and $4.6 \cdot 10^{-5}$, respectively.  Since
the leptons do not interact strongly, after they are produced, they
exit the dense system essentially unaffected carrying out crucial
information from the core of the system.  There is a good analogy in
understanding the processes in the center of the sun via neutrino
emission, since only neutrinos have a small enough interaction
cross-section to pass out of the sun's core largely unaffected.  In
the case of neutrinos, the more interesting physics of possible
neutrino oscillations complicates matters, but that is not a concern
in the case of our electron and muons measurements.

One other additional point of interest is that the apparent branching
ratio of the $\rho$ into pions and leptons should be modified as
observed by experiment.  The $\rho$ mesons that decay in medium into
pions are not reconstructed, but the ones decaying into electrons are.
If one can reconstruct the $\rho$ in both channels one can gauge the
number of lifetimes of the $\rho$ the dense medium survives for.

The lifetime of the $\rho$ meson in the rest frame of the plasma
depends on its gamma boost in this frame, and thus to study the
earliest stages, a measurement of low transverse momentum $\rho$
mesons is desirable.  If the $\rho$ decays at rest in the plasma
frame, the maximum transverse momentum for the electron or positron is
$\pt\, \approx 385$ \mevc.  These electrons are considered low \pt\
and present an experimental challenge to measure for two reasons.
First, there are a large number of low momentum charged pions created
in these collisions, that results in a charged pion to electron ratio
in this \pt\ range of $1000:1$.  Thus one needs detectors
that can cleanly identify electrons with good momentum resolution,
while rejecting the copiously produced pions.  The second challenge is
that most of the electrons come from pion Dalitz decays ($\pi^{0}
\rightarrow e^{+}e^{-}\gamma$), $\eta$ Dalitz decays ($\eta
\rightarrow e^{+}e^{-}\gamma$), and conversions of photons ($\gamma
\rightarrow e^{+}e^{-}$) mostly resulting from $\pi^{0}$ decays.
Ideally one wants to reject these other electrons to enhance
the signal contribution from the low mass vector mesons.  Conversions
are reduced by reducing the amount of material in the path of produced
photons.  This restriction is often at odds with the desire to have
substantial inner tracking detectors and these needs must be balanced.

The $\phi$(1020) meson spectral function is also sensitive to
in-medium chiral symmetry restoration; however its substantially
longer lifetime $\tau \approx 40~\mathrm{fm}/c$ means that most $\phi$ decays
occur outside the medium. However, measuring the $\phi$ in its various
decay modes (kaon pairs, electron pairs, muon pairs) remains an
interesting signal at low transverse momentum.

These low mass vector mesons also decay into muons pairs.  These muons
have low momentum and are an real experimental challenge to measure as
detailed later in these proceedings.

\subsection{Deconfinement}

There are many signatures that result from the deconfinement of color
charges over an extended volume, often referred to as the quark-gluon
plasma.  Two are detailed in this proceedings: (1) Suppression of
heavy quarkonium states and (2) Parton energy loss via gluon emission,
also referred to as jet quenching.

\subsubsection{Quarkonium Suppression}

\begin{figure}
    \centerline{\includegraphics[width=0.90\textwidth]{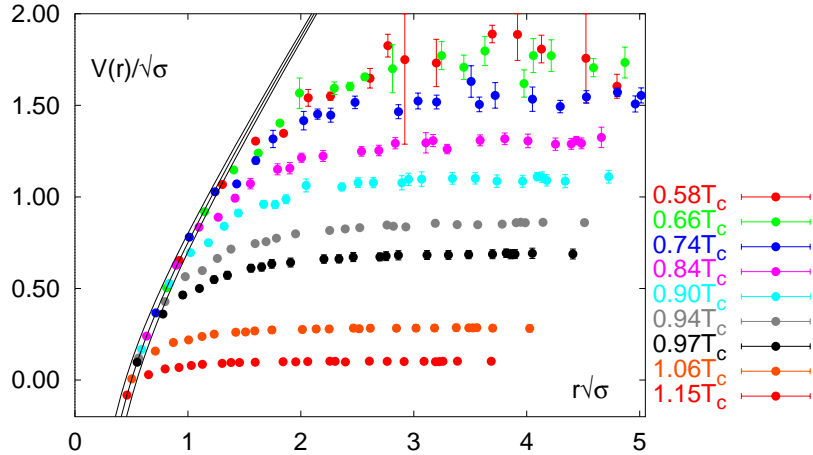}}
\caption{The QCD potential between heavy quarks in shown as a function
    of $r\sqrt{\sigma}$}
\label{karsch-confine}
\end{figure}

The suppression of heavy quarkonium states was originally proposed by
Matsui and Satz \cite{satz} in the late 1980's as a signature for color
deconfinement.  The Debye screening in a QED plasma is a reasonable
analogue for the scenario in our QCD plasma.  A charm-anticharm
($c\overline{c}$) quark pair produced via gluon fusion in the initial
phase of the heavy ion collision can form a $J/\psi$ if the pair has
low relative momentum.  The total production of such states in
proton-proton collisions relative to the total charm production is
less than a few percent.  If the Debye screening length is of order
the same size as the quarkonium state, then the pair is screened.  The
charm and anticharm quark scatter away from each other and, eventually
at the hadronization point, pair with surrounding light quarks and
antiquarks to form $D$ mesons.  This color screening is displayed in
recent lattice QCD calculations described at this workshop in terms of
a modification in the linear rise at large distances of the QCD
potential.  This change in the QCD potential as a function of
temperature is shown in Fig.~\ref{karsch-confine}.

There are a variety of heavy vector mesons with a large range in
binding energy (and associated hadronic size).  The $J/\psi$,
$\chi_{c}$, and $\psi'$ have binding energies of 0.64, 0.20, and 0.05
GeV respectively.  The $\Upsilon(1s)$, $\chi_{b}$, $\Upsilon(2s)$,
$\chi_{b}'$, and $\Upsilon(3s)$ have binding energies of 1.10, 0.67,
0.54, 0.31, 0.20 GeV respectively.  Since the suppression of these
states is determined by the relative plasma temperature and the
binding energy (or by the quarkonium hadronic size and the Debye
screening length), measuring the sequential disappearance of these
states acts as a QCD thermometer.  

The $J/\psi$(1s) state decays into almost anything hadronic with a
large branching ratio of 87.7\%.  However, the experimentally
accessible decay channels are 5.93\% to $e^{+}e^{-}$ and 5.88\% to
$\mu^{+}\mu^{-}$.  Similar decay channels are available for the
$\psi'$ and the $\Upsilon$ states.  The accessible decay channel of
the $\chi_{c}$ for heavy ion experiments is $\chi_{c} \rightarrow
\gamma + J/\psi$ with a branching ratio $6.6 \cdot 10^{-3}$.  Since
the decay $\gamma$ is quite soft, the $\chi_{c}$ represents an
experimental challenge.

Similar to the low mass vector mesons there is interest in these
states at low transverse momentum, where they reside in the plasma
state longer.  At rest the $J/\psi$ decays into electrons or muons
with a characteristic $\pt\, \approx 1.5$ GeV.  A rough rate estimate
(good to a factor of 2-3) is that the production of $J/\psi$ is
approximately $1 \cdot 10^{-4}$ per proton-proton collision.  In a
central Au-Au collision, there are of order 800 binary collisions,
yielding a $J/\psi$ rate of $8 \cdot 10^{-2}$.  The branching ratio to
electrons is 5.9\% and a typical experimental acceptance is 1\%,
yielding $4 \cdot 10^{-5}$ $J/\psi$ per Au-Au central collision, and
that is assuming no anomalous suppression!  Hence, one requires a
detector that measures either electron or muon pairs with a high
efficiency and a trigger and data acquisition system capable of
sampling events at the full RHIC design luminosity.  In particular, if
one wants to bin the data in terms of $x_{F}$, \pt, and collision
centrality, large statistics are a requirement.

There is a recent proposal for $J/\psi$ enhancement.  This scenario
assumes copious charm prodcution, and then at the hadronization stage
some charm and anticharm quarks may be close in both momentum and
configuration space.  These $c\overline{c}$ pairs may coalesce to form
$J/\psi$.  This late stage production would potentially mask any
suppression in the early stages from color screening.  There is an
easy test of this theory.  When RHIC runs at lower energies, for
example $\sqrt{s} = 60$ GeV, instead of the maximum energy of
$\sqrt{s} = 200$ GeV, the charm production is lower by a factor of
approximately three and the effect of recombination should be reduced
substantially.  In addition, we expect different \pt\ dependence of
$J/\psi$ production from original hard processes compared with late
stage $c\overline{c}$ coalescence.

\subsubsection{Parton Energy Loss}

An ideal experiment would be to contain the quark-gluon plasma, and
send well calibrated probes through it, and measure the resulting
transparency or opacity of the system.  There is no experimental way
of aiming a third beam of particle at the collision.  Therefore any
probes of the system must be generated in the collision itself.  These
probes must have calculable production rates in order to be considered
calibrated.  An excellent example of such a probe is a hard scattered
parton.  A parton traversing a color confined medium of hadrons sees a
relatively transparent system.  However, a parton passing through a
hot colored deconfined medium will lose substantial energy via gluon
radiation\cite{mueller,mueller2}.

The source of these partons is from hard scattering processes producing
back-to-back parton jets.  In a deconfined medium the parton will lose
energy before escaping the system and fragmenting into a jet cone of
hadrons.  The total energy of the initial parton jet is conserved
since eventually the radiated gluons will also hadronize.  It is
likely that the radiated gluons will have a larger angular dispersion
than the normally measured jet cone.  Thus one might be able to
measure a modification in the apparent jet shape.  
When the parton fragments into hadrons it has
less energy, and hence the fragmentation will result in a much reduced energy
for the leading hadron.  A measurement of high transverse
momentum hadrons ($\pi^{0}, \pi^{+/-}, K^{+/-}, h^{+/-}$) is a strong
indicator of the opacity of the medium.  An exciting additional
observable was mentioned at the workshop in the context of 
the high \pt\ spectra
of charm $D$ mesons.  There is a reduction in the induced gluon
radiation for charm and bottom quarks relative to light quarks due to
their slower velocity through the medium.  

\section{The RHIC Complex}

The scope of the RHIC program is to operate a colliding beam facility
which allows studies of phenomena in ultra-relativistic heavy-ion
collisions and in collisions of polarized protons.  The collider is
located in the northwest section of the Brookhaven National Laboratory
(BNL) in Upton, New York. Its construction began in 1991 and the
completion of the complex was accomplished in Spring 2000.

\begin{figure}[htb]
    \includegraphics[width=\textwidth]{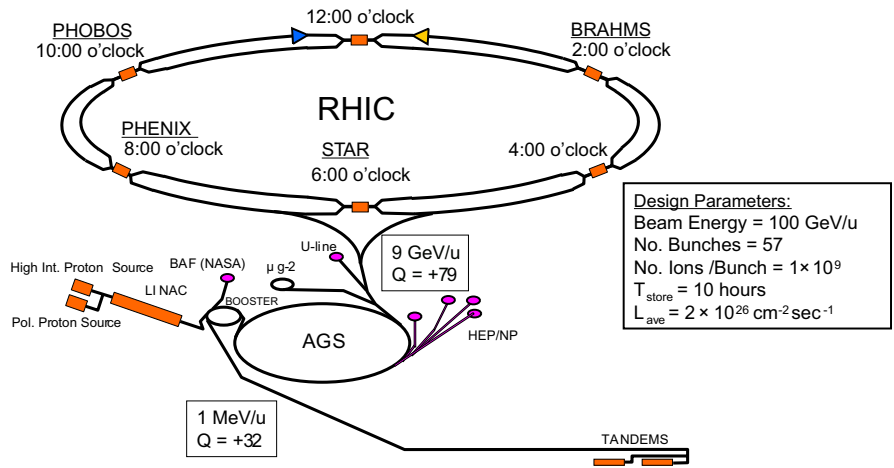}
\caption{The Relativistic Heavy Ion Collider (RHIC) accelerator
    complex at Brookhaven National Laboratory.  Nuclear beams are
    accelerated from the tandem Van de Graaff, through the transfer
    line into the AGS Booster and AGS prior to injection into RHIC.
    Details of the characteristics of proton and Au beams are also
    indicated after acceleration in each phase.}
\label{fig:rhicoverview}
\end{figure}

The collider, which consists of two concentric rings of 1740
super-\-con\-ducting magnets, was constructed in an already existing
ring tunnel of $\sim 3.8$ km circumference. This tunnel was originally
constructed for the proposed ISABELLE project. It offers an
extraordinary combination of energy, luminosity and polarization. A
schematic diagram of the whole RHIC complex, including the various
facilities used to produce and pre-accelerate the beams of particles
is displayed in Figure~\ref{fig:rhicoverview}.

\begin{figure}[hbt]
  \begin{center}
      \includegraphics[width=0.85\textwidth]{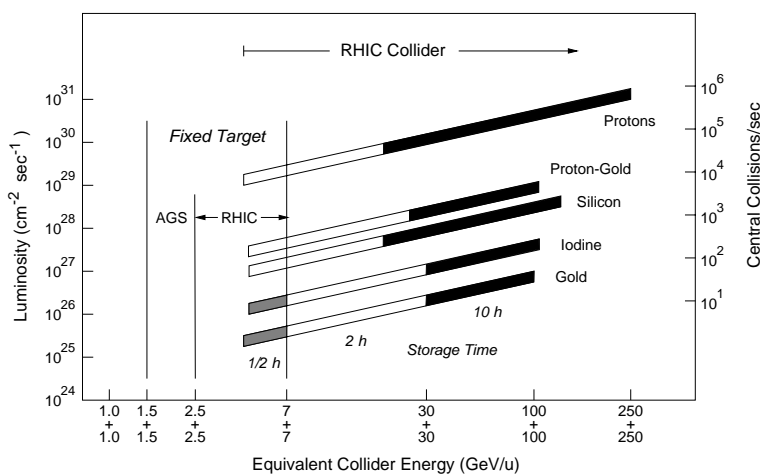}
    \caption{RHIC performance parameters.}
    \label{fig:rhicparameters}
  \end{center}
\end{figure}

RHIC is able to accelerate and store counter-rotating beams of ions
ranging from those of gold to protons at the top energy of 100
GeV/nucleon for gold and 250 GeV for protons.  The stored beam
lifetime for gold in the energy range of 30 to 100 GeV/nucleon is
expected to be approximately 10 hours.  The major performance
parameters are summarized in Figure~\ref{fig:rhicparameters}.


The layout of the tunnel and the magnet configuration allow the two
rings to intersect at six locations along their circumference.  The
top kinetic energy is 100+100 GeV/nucleon for gold ions. The
operational momentum increases with the charge-to-mass ratio,
resulting in kinetic energy of 125 GeV/nucleon for lighter ions and
250 GeV for protons. The collider is able to operate a wide range from
injection to top energies.  The collider is designed for a Au-Au
luminosity of about $2 \cdot 10^{26}$ cm$^{-2}$ s$^{-1}$ at top
energy. This design corresponds to approximately 1400 Au-Au minimum
bias collisions per second.
The luminosity is energy dependent and decreases approximately
proportionally as the operating energy decreases. For lighter ions it
is significantly higher reaching $\sim 1 \cdot 10^{31}$ cm$^{-2}$
s$^{-1}$ for pp collisions.  The collider allows collisions of beams
of equal ion species all the way down to pp and of unequal species
such as protons on gold ions.  Another unique aspect of RHIC is the
ability to collide beams of polarized protons (70-80\%) which allows
the measurement of the spin structure functions for the sea quarks and
gluons.

The first physics run at the Relativistic Heavy Ion Collider (RHIC)
took place in the Summer of 2000. For this run beam energies were kept
to a moderate 65 A\,GeV.  RHIC attained its goal of ten percent of
design luminosity by the end of its first run at the collision
center-of-mass energy of \sqrtsNN\ = 130 GeV.

In the following we describe briefly the various facilities, depicted
in Figure~\ref{fig:rhicoverview}, that are part of the large RHIC
complex:

\begin{description}
\item[Tandem Van de Graaff] Completed in 1970, the Tandem Van de
    Graaff facility was for many years the world's largest
    electrostatic accelerator facility. It can provide beams of more
    than 40 different types of ions ranging from hydrogen to uranium.
    The facility consists of two 15 MV electrostatic accelerators,
    each about 24 meters long, aligned end-to-end.In the Tandem the
    atoms are stripped of some of their electrons (e.g. Au to Q =
    +32) and accelerated to a kinetic energy of 1 MeV/nucleon.
    
\item[Heavy Ion Transfer Line (HITL)] To study heavy ion collisions at
    high energies, a 700 meter-long tunnel and beam transport system
    called the Heavy Ion Transport Line were completed in 1986,
    allowing the delivery of heavy ions from the Tandem to the Booster
    for further acceleration.  The HITL makes it possible for the
    Tandem to serve as the Relativistic Heavy Ion Collider's ions
    source.
    
\item[Linear Accelerator (Linac)] For the study of pp or pA collisions
    at the experiments, energetic protons are supplied by an Linear
    Accelerator (Linac).  The Brookhaven Linear Accelerator was
    designed and built in the late 1960's as a major upgrade to the
    Alternating Gradient Synchroton (AGS) complex.  The basic
    components of the Linac include ion sources, a radiofrequency
    quadrapole pre-injector, and nine accelerator radiofrequency
    cavities spanning the length of a 150 m tunnel.  The Linac is
    capable of producing up to a 35 milliampere proton beam at
    energies up to 200 MeV for injection into the AGS Booster.
    
\item[Booster] The Alternating Gradient Synchrotron Booster is less
    than one quarter the size of the AGS.  It is used to preaccelerate
    particles entering the AGS ring and plays an important role in the
    operation of the Relatavistic Heavy Ion Collider (RHIC) by
    accepting heavy ions from the Tandem Van de Graaff facility via
    the Heavy Ion Transfer Line (HITL) and protons from the Linac.  It
    then feeds them to the AGS for further acceleration and delivery
    to RHIC. After the installation of the HITL in 1986, the AGS was
    capable of accelerating ions up to silicon with its atomic mass of
    28.  However, due to its superior vacuum, the Booster makes it
    possible for the AGS to accelerate and deliver heavy ions up to
    gold with its atomic mass of 197.
                            
\item[AGS] Since 1960, the Alternating Gradient Synchrotron (AGS) has
    been one of the world's premiere particle accelerators and played
    a major role in the study of relativistic heavy ion collisions in
    the last decade.  The AGS name is derived from the concept of
    alternating gradient focusing, in which the field gradients of the
    accelerator's 240 magnets are successively alternated inward and
    outward, permitting particles to be propelled and focused in both
    the horizontal and vertical plane at the same time.  Among its
    other duties, the AGS is now used as an injector for the
    Relativistic Heavy Ion Collider. For RHIC operation the fully
    stripped ions are accelerated in the AGS to 9 GeV/nucleon before
    ejection.
                                                       
\item[ATR]The AGS sends the ions (or protons) down another beamline
    called the AGS-to-RHIC Transfer Line (ATR). At the end of this
    line, there's a "fork in the road", where sorting magnets separate
    the ion bunches.  From here, the counter-rotating beams circulate
    in the RHIC where they are collided at one of four intersecting
    points.
\end{description}

\section{Experimental Program}

\subsection{Letters of Intent}

In July 1991 there were a set of experimental Letters of Intent that were
put forward to an advisory committee.  The proposals are listed below, 
including the lead institution and in parenthesis the physics observable
focus.  

\begin{enumerate}
\item LBL-TPC (inclusive charged hadrons)
\item BNL-TPC (inclusive charged hadrons)
\item TOYKO-TALES (electron pairs, hadrons)
\item SUNY-SB (direct photons)
\item Columbia-OASIS (electron pairs, hadrons, high \pt)
\item ORNL Di-Muon (muon pairs)
\item BNL Forward Angle Spectrometer (hadrons at large rapidity)
\item MIT MARS (hadrons and particle correlations)
\end{enumerate}

The experiments span the range of hadronic, leptonic and photonic
capabilities to cover the broad spectrum of physics topics listed above.
At the time, only the LBL-TPC proposal was approved and became the STAR
experiment.  Eventually the MARS proposal evolved into the PHOBOS experiment 
(note that Phobos is a moon of the planet Mars) and PHENIX (should be spelled
Phoenix) rose from the ashes of OASIS, Di-Muon, TALES and the other lepton
focussed experiments.

Eventually there were four approved experiments which have now been 
constructed and operated during the first year of RHIC running.  
BRAHMS, PHENIX, PHOBOS and STAR are briefly described below.
These experiments have
various approaches to study the deconfinement phase transition to the
quark gluon plasma. The STAR experiment \cite{STAR} concentrates on
measurements of hadron production over a large solid angle in order to
measure single- and multi-particle spectra and to study global
observables on an event-by-event basis.  The PHENIX experiment
\cite{PHENIX} focuses on measurements of lepton and photon production
and has the capability of measuring hadrons in a limited range of
azimuth and pseudo-rapidity.  The two smaller experiments BRAHMS (a
forward and mid-rapidity hadron spectrometer) \cite{BRAHMS} and PHOBOS
(a compact multiparticle spectrometer) \cite{PHOBOS} focus on single-
and multi-particle spectra.  The collaborations, which have
constructed these detector systems and which will exploit their
physics capabilities, consist of approximately 900 scientists from
over 80 institutions internationally.  In addition to colliding heavy
ion beams, RHIC will collide polarized protons to study the spin
content of the proton \cite{Spin}.  STAR and PHENIX are actively
involved in the spin physics program planned for RHIC.

\subsection{BRAHMS}
The \textbf{BR}oad \textbf{R}ange \textbf{H}adron \textbf{M}agnetic
\textbf{S}pectrometer BRAHMS experiment is designed to measure and
identify charged hadrons ($\pi^\pm$, K$^\pm$, $^{(}
\overline{\mathrm{p}} ^{)}$) over a wide range of rapidity and
transverse momentum for all beams and energies available at RHIC.
Because the conditions and thus the detector requirements at
mid-rapidity and forward angles are different, the experiment uses two
movable spectrometers for the two regions.

\begin{figure}[htb]
    \begin{center}
        \includegraphics[width=0.7\textwidth]{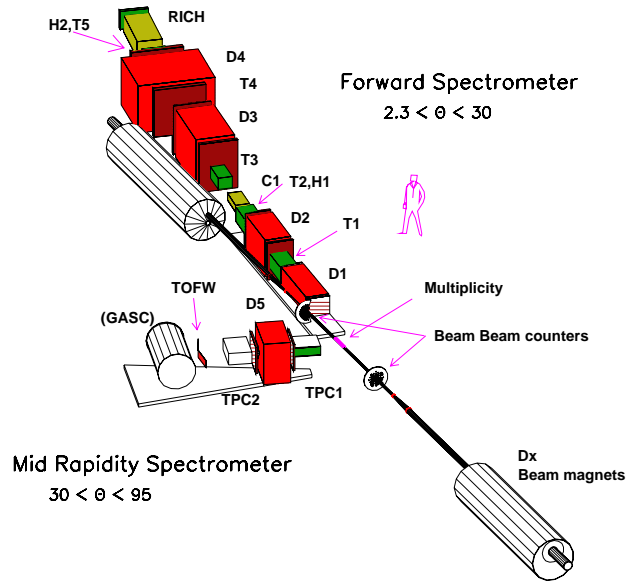}
    \end{center}
\caption{Layout of the BRAHMS detector.}
\label{fig:brahms}
\end{figure}

As shown in Figure~\ref{fig:brahms}, there is a mid-rapidity spectrometer
to cover the pseudo-rapidity range $0 \le \eta \le 1.3$ and a
forward spectrometer to cover $1.3 \le \eta \le 4.0$.  The latter
employs four dipole magnets, three time projection chambers
(TPC), and drift chambers. Particle
identification is achieved with time-of-flight hodoscopes, a
threshold Cherenkov counter, and one ring-imaging Cherenkov counter
(RICH). The solid angle acceptance of the forward arm is 0.8 mstr.
The mid-rapidity spectrometer has been designed for charged particle
measurements for p $\le 5\, \mathrm{GeV}/c$. The spectrometer has two TPCs for
tracking, a magnet for momentum measurement, and a time-of-flight
wall and segmented gas Cherenkov counter (GASC) for particle
identification. It has a solid angle acceptance of 7 mstr. A set of
beam counters and a silicon multiplicity array provide the experiment
with trigger information and vertex determination.

\subsection{PHOBOS}

\begin{figure}[bht]
\includegraphics[width=\textwidth]{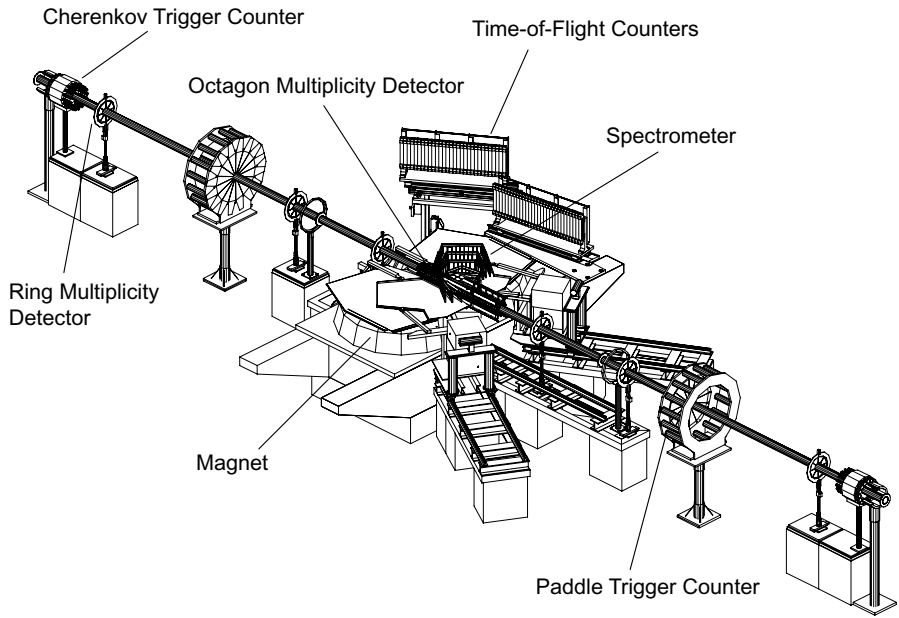}
\caption{PHOBOS detector setup for the 2000 running period.}
\label{fig:phobos}
\end{figure}

The PHOBOS detector is designed to detect as many of the produced
particles as possible and to allow a momentum measurement down to very
low \pt.  The setup consists of two parts: a multiplicity
detector covering almost the entire pseudo-rapidity range of the
produced particles and a two arm spectrometer at mid-rapidity. Figure
\ref{fig:phobos} shows the detector, including the spectrometer arms,
the multiplicity and vertex array, and the lower half of the magnet.

One aspect of the design is that all detectors are produced using a
common technology, namely as silicon pad or strip detectors.  The
multiplicity detector covers the range $-5.4 < \eta < 5.4$, measuring
total charged multiplicity dN$_{\mathrm{ch}}$/d$\eta$ over almost the
entire phase space. For approximately 1\% of the produced particles,
information on momentum and particle identification will be provided
by a two arm spectrometer located on either side of the interaction
volume (only one arm was installed for the 2000 run). Each arm covers
about 0.4 rad in azimuth and one unit of pseudo-rapidity in the range
$0 < \eta < 2$, depending on the interaction vertex, allowing the
measurement of p$_\perp$ down to 40 MeV/$c$. Both detectors are capable
of handling the 600 Hz minimum bias rate expected for all collisions
at the nominal luminosity.

\subsection{PHENIX}
\begin{figure}[htb]
    \centerline{\includegraphics[width=0.77\textwidth]{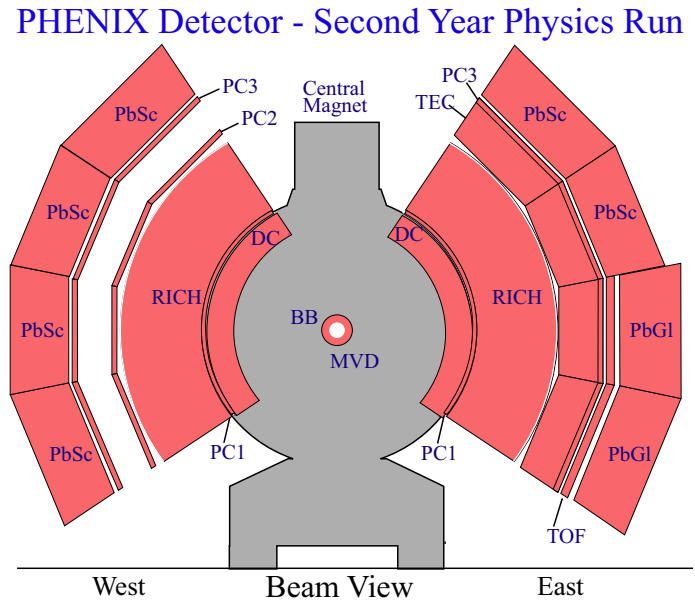}}
\caption{Shown is a beam view of the PHENIX two central spectrometer
    arms.  Their is a axial field magnet in the middle.  The detectors
    from the inner radius out are the multiplicity and vertex detector
    (MVD), beam-beam counters (BBC), drift chambers (DC), pad chambers
    (PC1-3), ring imaging cherenkov counter (RICH), time-expansion
    chamber (TEC), time-of-flight (TOF), and a Lead Glass and Lead
    Scintillator electro-magnetic calorimeter (PbSc, PbGl).}
\label{phenix-detector}
\end{figure}

The PHENIX experiment is specifically designed to measure
electrons, muons, hadrons and photons. The experiment is capable
of handling high event rates, up to ten times RHIC design
luminosity, in order to sample rare signals such as the $J/\psi$
decaying into muons and electrons, high transverse momentum
$\pi^{0}$'s, direct photons, and others.  The detector consists of
four spectrometer arms.  Two central arms have a small angular
coverage around central rapidity and consist of a silicon vertex
detector, drift chamber, pixel pad chamber, ring imaging Cerenkov
counter, a time-expansion chamber, time-of-flight and an
electromagnetic calorimeter.  These detectors allow for electron
identification over a broad range of momenta in order to measure
both low mass and high mass vector mesons. Two forward
spectrometers are used for the detection of muons. They employ
cathode strip chambers in a magnetic field and interleaved layers
of Iarocci tubes and steel for muon identification and triggering.
The overall layout of the PHENIX detector is shown in
Figure~\ref{phenix-detector}.  

\begin{figure}[htb]
\centerline{\includegraphics[width=0.90\textwidth]{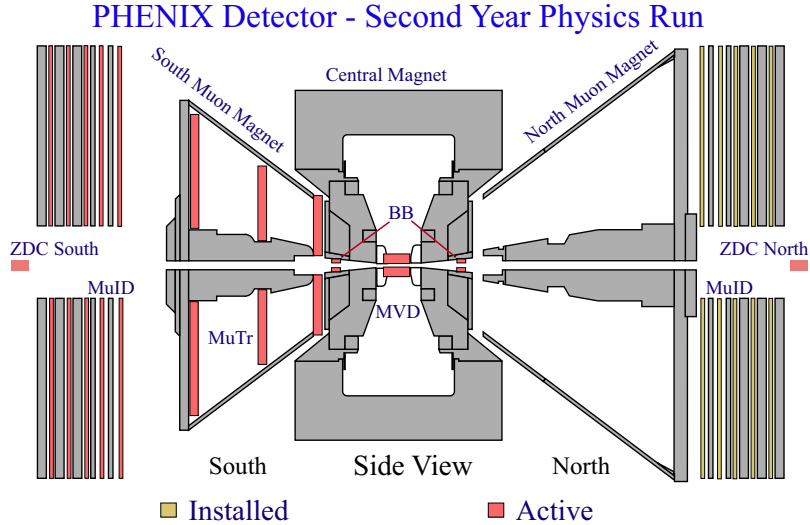}}
\caption{Shown is a side view of the PHENIX detector including the two 
central spectrometer arms and the two muon spectrometers.  The muons
systems consist of cathode strip chamber muon trackers (MuTr) and
muon identifiers (MuID) interleaved with layers of steel.}
\label{phenix_magnet}
\end{figure}

One key feature of the PHENIX detector is the high rate capability of
the data acquisition system (DAQ) and multi-level trigger
architecture.  These allow PHENIX to sample physics from RHIC
collisions above the design luminosity of the machine.  This high rate
is crucial for studying rare leptonic, photonic and high \pt\
processes.

\subsection{STAR}

\begin{figure}[htb]
\includegraphics[width=\textwidth]{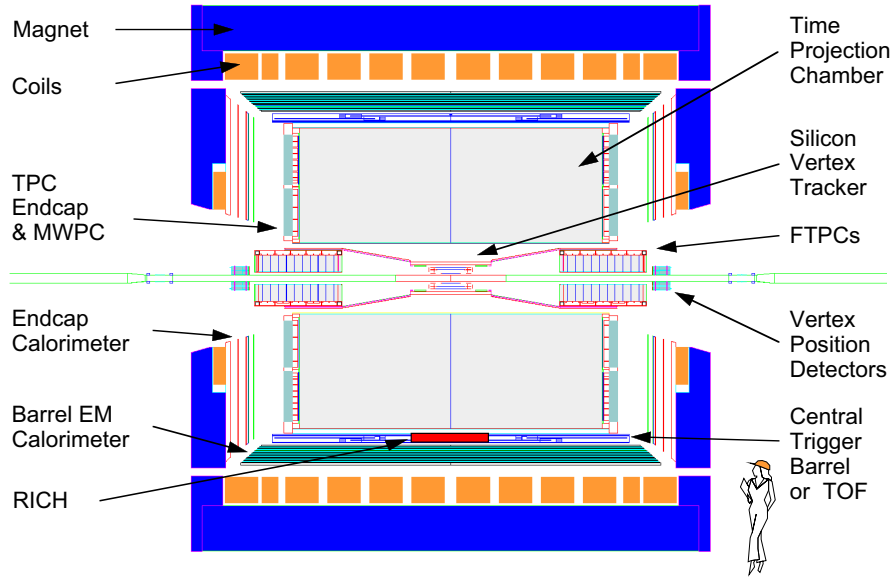}
\label{fig:star}
\caption{Schematic view of the STAR detector.}
\end{figure}

The \textbf{S}olenoidal \textbf{T}racker \textbf{A}t \textbf{R}HIC
(STAR) is a large acceptance detector capable of tracking charged
particles and measuring their momenta in the expected high
multiplicity environment.  It is also designed for the measurement and
correlations of global observables on an event-by-event basis and the
study of hard parton scattering processes.  The layout of the STAR
experiment is shown in Figure~\ref{fig:star}.  The initial
configuration of STAR in 2000 consists of a large time projection
chamber (TPC) covering $|\eta| < 2$, a ring imaging Cherenkov detector
covering $|\eta| < 0.3$ and $\Delta{\phi} = 0.1\pi$, and trigger
detectors inside a solenoid\-al magnet with 0.25 T magnetic field. The
solenoid provides a uniform magnetic field of maximum strength 0.5 T
for tracking, momentum analysis and particle identification via
ionization energy loss measurements in the TPC.  Measurements in the
TPC were carried out at mid-rapidity with full azimuthal coverage
($\Delta{\phi} = 2\pi$) and symmetry. A total of 1M minimum bias and
1M central events were recorded during the summer run 2000.

Additional tracking detectors will be added for the run in 2001.
These are a silicon vertex tracker (SVT) covering $|\eta| < 1$ and two
Forward TPCs (FTPC) covering $2.5 < |\eta| < 4$.  The electromagnetic
calorimeter (EMC) will reach approximately 20\% of its eventual $-1 <
\eta < 2$ and $\Delta{\phi} = 2\pi$ coverage and will allow the
measurement of high transverse momentum photons and particles.  The
endcap EMC will be constructed and installed over the next 2 -- 3
years.

\subsection{RHIC Spin Program}
The design of both the STAR and PHENIX experiments includes a
polarized proton program to conduct studies of the spin structure of
the proton.  Critical to this measurement is the identification of
high transverse momentum photons and leptons. The STAR experiment is
phasing in an electromagnetic calorimeter that will be crucial for
such observations.  In particular the RHIC experiments are ideally
suited for measuring the gluon contribution of the proton spin.

\section{Experimental Techniques}

\subsection{Tracking Charged Particles (The STAR Example)}

The STAR experiment aims at the observation of hadronic observables
and their correlations, global observables on an event-by-event base,
and the measurement of hard scattering processes.
The physics goals dictate the design of an experiment. Given the physics
directions it is easy to summarize the general requirements:
\begin{itemize}
\item Soft physics (100 MeV/$c < \pt\,< 1.5$ \gevc)
    \begin{itemize}
    \item detection of as many charged particles as possible with high
        efficiency to provide high statistics for event-by-event
        observables and fluctuation studies
    \item $2\pi$ \textit{continuous} azimuthal coverage for reliable event
        characterization
    \item high tracking efficiency as close to the vertex as possible
        to contain the size of the experiment
    \item adequate track length for tracking, momentum measurement and
        particle identification for a majority of particles
    \item good two-track resolution providing a momentum difference
        resolution od a few \mevc\ for HBT studies
    \item accurate determination of secondary vertices for detecting
        strange particles ($\Lambda, \Xi, \Omega$)
    \end{itemize}
\item Hard physics ($> 1.5$ \gevc\ and jets)
    \begin{itemize}
    \item large uniform acceptance to maximize rates and minimize edge
        effects in jet reconstruction
    \item accurate determination of the primary vertex in order to
        achieve high momentum resolution for primary particles
    \item electromagnetic calorimetry combined with tracking and good
        momentum resolution up to $\pt\,=$ 12 \gevc\ to trigger on jets
    \item segmentation of electromagnetic calorimeters which is
        considerable finer than the typical jet size,
        \textit{i.e.}~jet radius $r = \sqrt{d\eta^2 + d\phi^2} \sim 1$.
     \end{itemize}
\end{itemize}
The challenge is to find a detector concept which meets all these
requirements with minimal costs. The detector of choice to solve the
main tracking tasks was a large Time Projection Chamber (TPC) operated
in a homogeneous magnetic field for continuous tracking, good momentum
resolution and particle identification (PID) for tracks below 1 \gevc.
The requirements not met by the TPC needed to be covered by more
specialized detectors such as a large acceptance electromagnetic
calorimeter (hard processes), two Forward-TPCs (coverage of $\eta >
1.7$), and an inner silicon vertex tracker (better primary and
secondary vertex measurement).  For particle identification of
high-\pt\ particles detectors a small Ring Imaging Detector (RICH) and
a Time-of-Flight (ToF) patch were added. Both detectors are not usable
for event-by-event physics due to their small acceptance but allow to
extend the PID capabilities for \textit{inclusive} distributions.

The need for an homogeneous field along the beam direction puts an
stringent constraint on the design of the whole experiment. Only large
so\-le\-no\-id\-al magnets are able to provide uniform fields of considerable
strength (0.5 T).  To keep down the costs the magnet cannot be too
large which limits significantly the amount of ``real estate''
(\textit{i.e.}~detectors) it can house inside for tracking, PID, and
calorimetry.  The final magnet has coils with an inner radius of 2.32
m and a yoke radius of 2.87 m.  The total length is 6.9 m. Note, that
this concept is very different from the design of PHENIX where the
axial-field magnet does not, or only weakly, constrain the dimensions
of the required detectors.
\begin{figure}[htb]
    \begin{center}
        \includegraphics[width=0.78\textwidth]{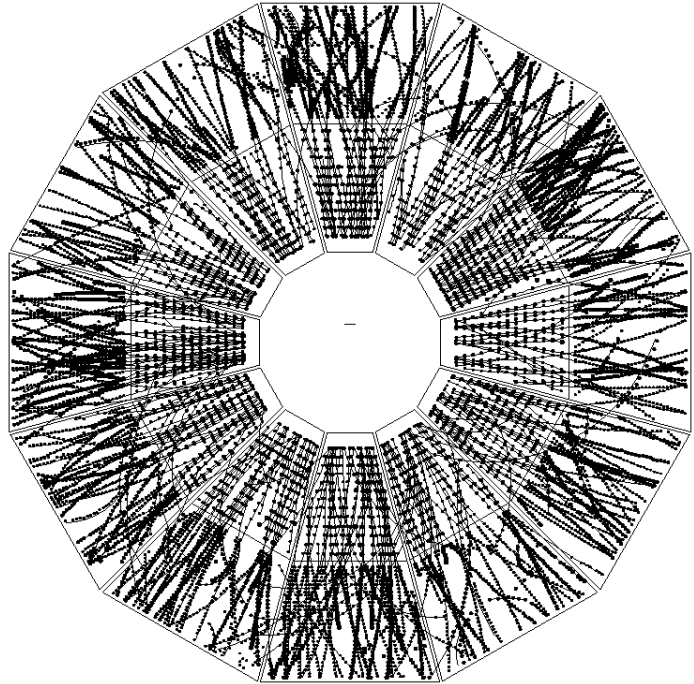}
\end{center}
\caption{Peripheral Au+Au event recorded in the STAR TPC. Shown is
    the projection of all hits (points) and reconstructed tracks
    (solid lines) in the event onto the xy plane perpendicular to the
    beamline.}
    \label{fig:stareventend}
\end{figure}

\begin{figure}[htb]
    \begin{center}
        \includegraphics[width=0.73\textwidth]{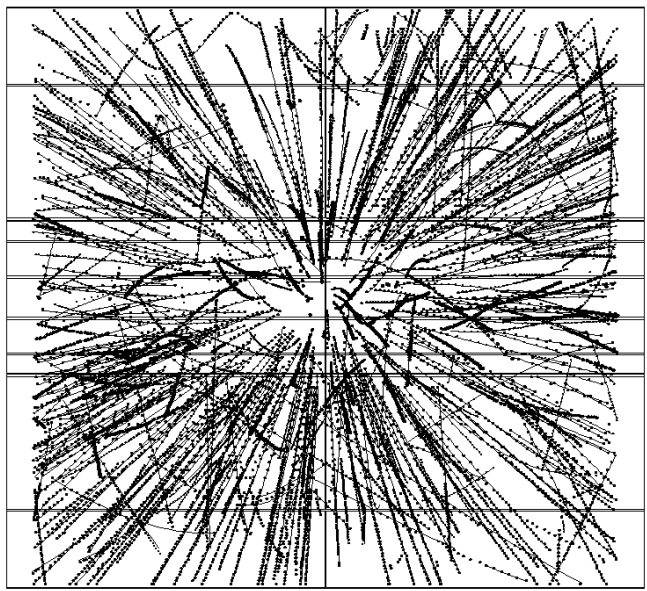}
\end{center}
\caption{Same event as shown in Fig.~\ref{fig:stareventend}
        but viewed from the side. The beamline runs horizontally from the
        left to the right.}
    \label{fig:stareventside}
\end{figure}

In the following we focus on the TPC, STARs main tracking detector,
that essentially performs the role of a 3D camera with around 70
million pixel resolution.  The TPC is divided into two longitudinal
drift regions, each 2.1 m long.  Electrons created from track
ionization drift in the longitudinal direction, along the TPC electric
field lines, to the end-caps of the TPC.  Each end-cap is instrumented
with 70,000 pads. Each pad reads out 512 time samples.  The position
of the ionization charge in the readout plane provides the x and y
coordinates of a space point along the particle trajectory while the
arrival \textit{time} of the charge allows to determine the original z
position.  The ionization pattern (chain of hits) of a traversing
charged particle curved in the magnetic field allows the complete
reconstruction of the particle trajectory and its 3-momentum.
Fig.~\ref{fig:stareventend} shows the xy-projection (front view) of a
low multiplicity event recorded in the STAR TPC. Each point represents
one reconstructed hit, i.e. the local ionization charge created by one
particle. The lines represent the reconstructed trajectories of the
particles.  Together with the timing information it is possible to
also reconstruct the z-position of the hits and such the polar angle
of the tracks as shown in Fig.~\ref{fig:stareventside}.  Sophisticated
pattern recognition programs perform the task of reconstructing the
particle trajectory using the measured positions of the measured hits.
This procedure is commonly referred to as ``tracking''.  In the
following we discuss the parameterization used to describe a track in
the STAR TPC (and in any other detector with an homogeneous solenoidal
field).

\subsubsection{Track Parameterization Momentum Determination}

The trajectory of a charged particle in a static uniform magnetic
field with $\vec{B} = (0, 0, B_z)$ is a helix. In principle five
parameters are needed to define a helix. From the various
possible parameterizations we describe here the version which is most
suited for the geometry of a collider experiment and therefore used
in STAR.

\begin{figure}[thb]
    \begin{center}
        \includegraphics[width=0.49\textwidth]{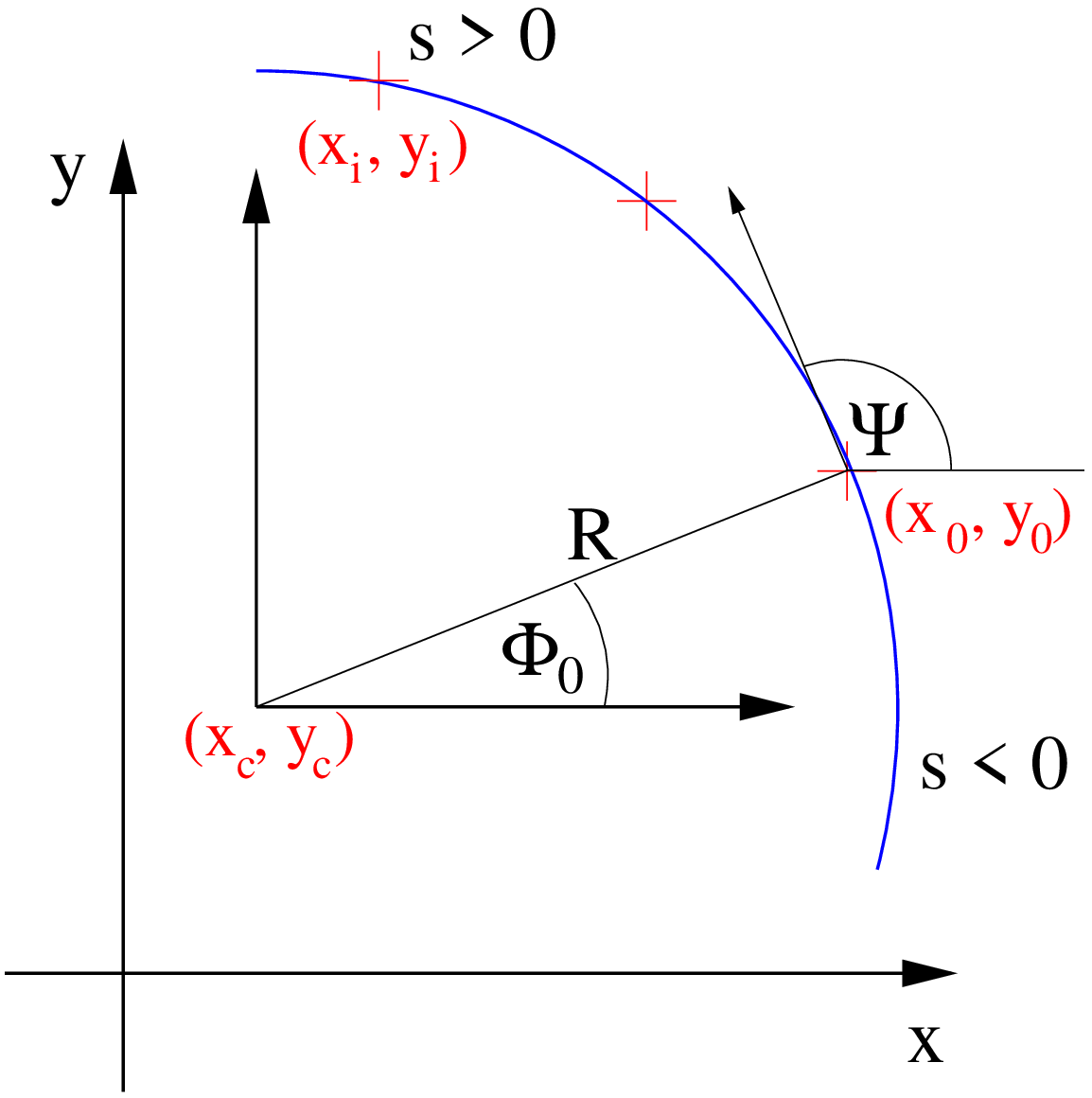}
        \includegraphics[width=0.49\textwidth]{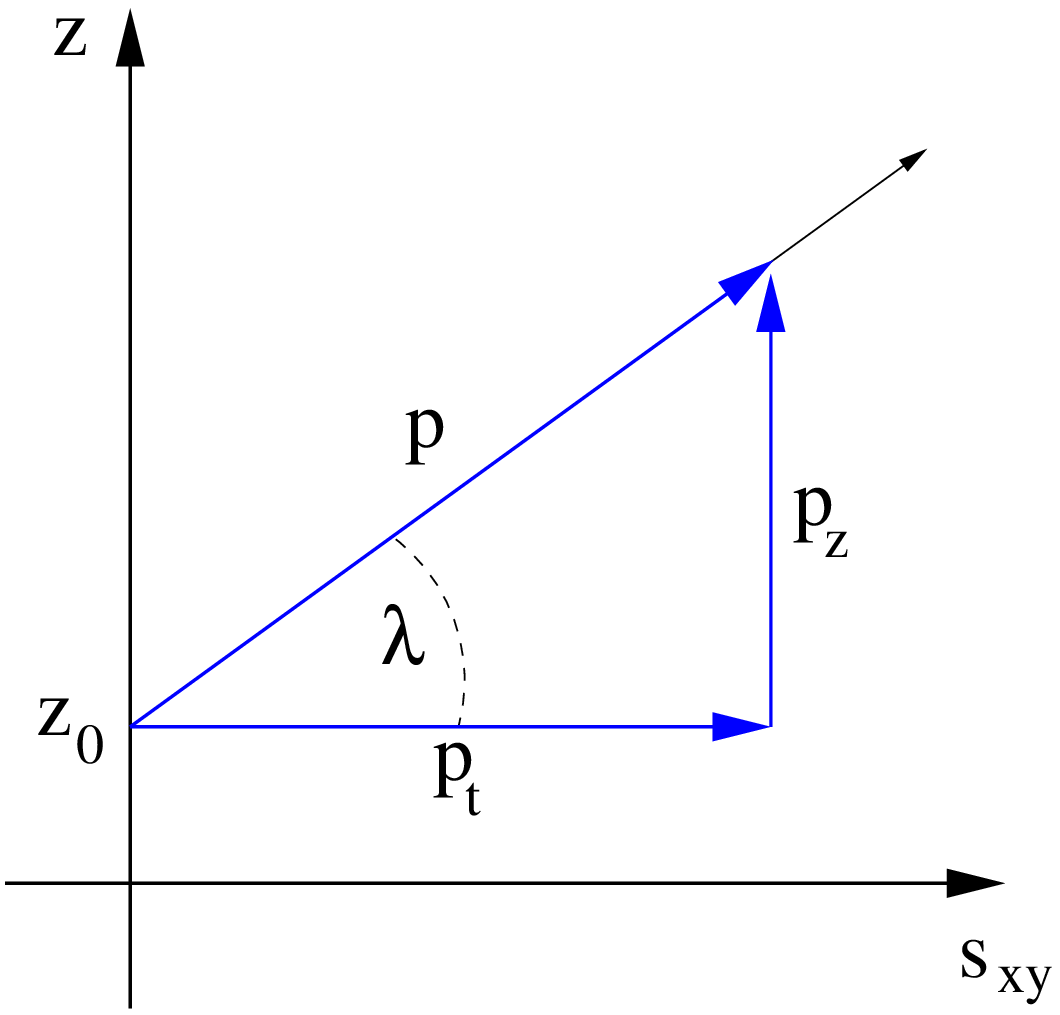}
\end{center}
\caption{Helix parameterization: shown on the left is the projection of a helix
    on the $xy$ plane.
    The crosses mark possible data points. The right plot
    depicts the projection of a helix on the $sz$ plane. For the
    meaning of the various parameters see text.}
  \label{fig:helix}
\end{figure}

This parameterization describes the helix in Cartesian coordinates,
where $x, y$ and $z$ are expressed as functions of the track length
$s$.

\begin{eqnarray}
    x(s) & = & x_0 + \frac{1}{\kappa} [\cos(\Phi_0 + h\ s\ \kappa\ \cos\lambda) - \cos\Phi_0] \label{eq:xs} \\
    y(s) & = & y_0 + \frac{1}{\kappa} [\sin(\Phi_0 + h\ s\ \kappa\ \cos\lambda) - \sin\Phi_0] \label{eq:ys} \\
    z(s) & = & z_0 + s\ \sin\lambda \label{eq:zs}
\end{eqnarray}
where:
$\mathbf{s}$ is the path length along the helix\\
$\mathbf{x_0, y_0, z_0}$ is the starting point at $s = s_0 = 0$\\
$\mathbf{\lambda}$ is the dip angle\\
$\mathbf{\kappa}$ is the curvature, i.e.~$\kappa = 1/R$\\
$\mathbf{B}$ is the z component of the homogeneous magnetic field ($B = (0, 0, B_z)$)\\
$\mathbf{q}$ is charge of the particle in units of positron charge\\
$\mathbf{h}$ is the sense of rotation of the projected helix in the $xy$-plane, \textit{i.e.}~ $h = -\mathrm{sign}(q B) = \pm 1$\\
$\mathbf{\Phi_0}$ is the azimuth angle of the starting point (in cylindrical coordinates) with respect to the helix axis ($\Phi_0 = \Psi - h \pi/2$)\index{phase}\\
$\mathbf{\Psi}$ is the $\arctan(\mathrm{d}y/\mathrm{d}x)_{s = 0}$, \textit{i.e.}~the azimuthal angle of the track direction at the starting point.\\
The meaning of the different parameters is visualized in Fig.~\ref{fig:helix}.

The circle fit in the $xy$-plane gives the center of the fitted circle $(x_c, y_c)$
and the curvature $\kappa = 1/R$ while the linear fit gives $z_0$ and $\tan \lambda$.
The phase of the helix (see Fig.~\ref{fig:helix}) is defined as follows:
\begin{eqnarray}
  \Phi_0 = \arctan \left( \frac{y_0 - y_c}{x_0 - x_c} \right)
\end{eqnarray}
The reference point $(x_0, y_0)$ is then calculated as follows:
\begin{eqnarray}
  x_0 & = & x_c + \frac{\cos \Phi_0}{\kappa} \\
  y_0 & = & y_c + \frac{\sin \Phi_0}{\kappa}
\end{eqnarray}
and the helix parameters can be evaluated as:
\begin{eqnarray}
  \Psi & = & \Phi_0 + h \pi / 2 \\
  \pt & = & c\ q\ B / \kappa \\
  p_z & = & \pt\, \tan \lambda \\
  p & = & \sqrt{p^2_\perp + p^2_z}
\end{eqnarray}
where $\kappa$ is the curvature in [m$^{-1}$], $B$ the value of the
magnetic field in [Tesla], $c$ the speed of light in [m/ns] ($\approx
0.3$) and \pt\ and p$_z$ are the transverse and longitudinal
momentum in [\gevc].

Once the track momenta and parameters are determined, one can use the
above parameterization to extrapolate the TPC tracks to other
detectors, e.g.~outwards to the electromagnetic calorimeter or RICH
detector, or inwards to the silicon vertex tracker (SVT).  Once the
referring hits are found they are added to the track and the parameters are
re-evaluated.  The charges of all hits along a track can be used to
calculate its energy loss ($dE/dx$) and together with its known momentum
can be used to determine the particle mass.
\subsection{Electron Identification (the PHENIX example)}

As we have seen in the physics motivation section, there are many
reasons to measure electrons over a broad range in transverse
momentum.  In the original Letters of Intent there were many different
proposals for detector technologies for measuring electrons, and many
of these capabilities were combined into the PHENIX experiment.  The
PHENIX experiment has two central spectrometer arms each covering 90
degrees in $\phi$ and $|\eta| < 0.35$.

\begin{figure}
    \centerline{\includegraphics[width=20pc]{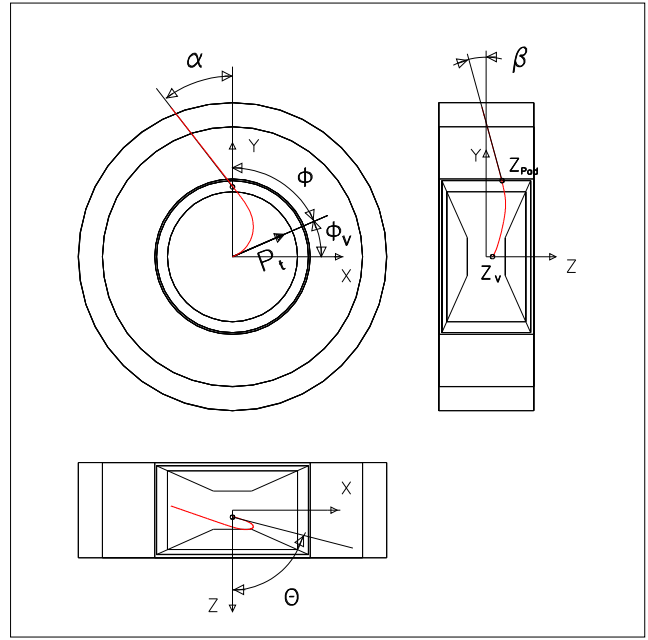}}
    \caption{A schematic diagram of the PHENIX central spectrometer
             magnetic reconstruction technique is shown.}
    \label{phenix:magnet2}
\end{figure}

Electrons and also charged hadrons that are produced near mid-rapidity
are bent in an axial magnetic field over a radial distance of
approximately two meters after which the aperture is relatively field
free.  There is a large aperture drift chamber that measures the
projective trajectory of the charged particle tracks in the field free
region with multiple wire layers oriented in the x direction (giving
maximum resolution in the bend plane of the field) and also in u and v
direction (stereo planes for pattern recognition).  The drift chamber
is augmented by a series of moderate resolution pad chambers, which
yield three dimensional space point along the particle's track and aid
significantly in pattern recognition and non-vertex background
rejection.  As shown in Figure~\ref{phenix:magnet2} the particle track
is characterized by an angle ($\alpha$) in the bend plane of the
magnetic field.  From this vector and the assumption that the track
originated at the Au-Au collision vertex, the rigidity of the track is
determined.  The rigidity is the momentum divided by the particle
charge, and with the assumption of a $Z=\pm 1$ particle, the momentum
is known.

Typically the momentum resolution has two components as shown below.
\begin{equation}
\delta p/p = (\approx 1\%) + (\approx 1\%) \times p \rm{[\gevc]}
\end{equation}
The first term, typically of order 1\% is due to the multiple
scattering of the charge particle in material before and in the
tracking devices.  The second term which scales with the particle's
momentum is related to the finite spatial resolution of the detector.
The PHENIX detector has been designed with a minimum of inner region
material including a Beryllium beam pipe, which has a low Z value to
minimize multiple scattering while maintaining integrity for the
vacuum.  All four experiments at RHIC have Be beam-pipes in the
interaction region.  The PHENIX design resolution from the drift
chamber is of order 0.5\% at $\pt\, \approx 0.2$ GeV and increases
linearly with \pt\ above 0.7 GeV.  There is a great deal of work
involved in calibrations and wire alignement to achieve these
resolution values.

It is notable that PHENIX uses an axial field magnet for the central
spectrometers as opposed to a solenoidal field.  There are substantial
advantages and disadvantages to this choice.  One major consideration
is that if the entire experiment is contained inside a large solenoid,
as in the STAR configuration, the magnetic field is present throughout
the volume.  This is advantageous for the large volume Time Projection
Chamber (TPC) of STAR where they can observe the curvature of the
charged particle tracks.  However, the cost of these magnets increases
steeply as one increased the desired outer radius.  PHENIX for the
purposes of hadron particle identification wanted to have a
Time-of-Flight scintillator wall approximately 5 meters from the
interaction vertex and for electron and photon identification a good
energy resolution and high granularity electromagnetic calorimeter
behind that.  In keeping within the experimental budget it would not
have been possible to have a solenoidal magnet with an outer radius
greater than 5.5 meters.  The ALICE experiment being built for the
CERN-LHC program is fortunate to be able to re-use the very large
solenoid from the L3 experiment at LEP.  One disadvantage of the
PHENIX choice is that the pole tips of the axial field magnet are
close to the interaction point and create substantial ``shine'',
particles scattering off the poles into the detector aperture.  The
overall choice for the field configuration is an important starting
point for many detector designs.

Now that we have characterized the particle's momentum vector, we must
discriminate all of the charged pions from our interesting electrons.
The first detector in PHENIX that is employed is a Ring Imaging
Cherenkov detector (RICH).  It is a large gas volume detector with a
thin mirror plane for reflecting Cherenkov light onto an array of
photo-multiplier tubes (PMT) that are situated off to the side of the
spectrometer acceptance. The radiator gas used is either ethane with
an index of refraction n=1.00082 or methane with n=1.00044.  Requiring
more than three PMT hits yields almost 100\% efficiency for electrons
and rejects pions with $\pt\, \approx < 4$ GeV at the level of $<
10^{3}$ in a single track environment.  The detector, requiring only
three PMT's, does not reconstruct a ring radius for further particle
characterization, but rather is used as a threshold detector only.

In a multiple track collisions (remember of order 5000 charged
particles are being produced) one can incorrectly match a track to a
RICH signal.  Further electron identification is provided by the
electromagnetic calorimeter (EMCal).  The calorimeter is composed of
both Lead Scintillator (PbSc) and Lead Glass (PbGl) modules, with the
later being originally used in the WA98 experiment at the CERN-SPS
heavy ion program.  The calorimeter has a radiation length of $\approx
18 X_{0}$ and $\approx 16 X_{0}$ for the PbSc and PbGl respectively,
but does not fully interact and contain hadronic showers.  Thus, an
electron or photon incident on the calorimeter deposits most of its
energy, while a hadron (eg. charged pion) has a large probability to
pass through the module depositing a small $dE/dx$ minimum ionizing
radiation energy.  Even when the pion suffers an inelastic collision
in the calorimeter, only a fraction of its energy is contained and
measured in the PMT at the back of the module.  The excellent energy
resolution $\approx 5-8\%/\sqrt{E(\mathrm{GeV})}$ and high granularity give
precise electron identification and pion rejection by requiring the
energy match the measured momentum ($E/p \approx
1$).  However, the background rejection degrades when the particle
momentum is low and hadronic showers have a higher probability to
match the measured momentum.

Therefore, an additional detector is necessary to help identify low
\pt\ electrons for the crucial low mass vector meson physics.  For
this purpose, PHENIX uses a Time Expansion Chamber (TEC) that samples
energy loss in a gas radiator.  The TEC determines the particle
species using $dE/dx$ information.  It has a rejection of $e/\pi
\approx 5\%$ for particles with a momentum p = 500 \mevc\ with P10 gas
and 2\% with Xe gas, which is much more expensive.  It is the
combination of all these detectors that allow for efficient electron
measurements with a minimum of pion contamination.  Thus, the
challenge of electron identification over a broad range in \pt\ is
met.

\subsection{Muon Identification (the PHENIX example)}

Direct muons and hadrons decaying into muons require a rather
different experimental approach to measuring electrons, photons and
other hadrons.  Muons interact with a low cross section in material,
and are easiest to identify by placing steel or other material in the
particle path and removing all other particles through interaction.
Then a detector placed after the steel should measure a clean muon
sample.

However, the muons must have a large enough energy to penetrate the
steel without stopping due to ionization energy loss $dE/dx$.  In the
central rapidity region at a collider, the muons from low \pt\
$J/\psi$ decays make it impossible to have enough steel to range out
other hadrons effectively, while allowing the muons to pass through.
In particle experiments that focus on high
\pt\ muons, the detector
can consist of a similar one to that described above for electrons.
After the calorimeter, one can have some steel absorber (often the
return in the magnet steel) and then have a muon identifier detector
outside of that.  This design does not work well for heavy ion
physics.  First, it restricts the measurable \pt\ range at too high
a value and also has a large background from low \pt\ pions
decaying into muons.

Another option is to measure muons are forward rapidity where they
have a substantial momentum in the longitudinal direction.  PHENIX
measures muons in the forward and backward pseudo-rapidity regions (at
angles of 10-35 degrees from the beam line).  The detector consists of
a brass and steel absorber to range out hadrons followed by a cathode
strip chamber muon tracking device that measures the particle's bend
in a magnetic field.  There is a delicate balance in the amount and
type of absorber material used.  Too much material and the multiple
scattering reduces the resolution which is important for cleanly
separating states such as $J/\psi$ and $\psi'$, and too little
material in which case the particle occupancy in the tracking device
is too large.  After the muon tracking device there is more steel
absorber with Iarocci tube muon identifiers interspersed.  The
coverage is from $1.1 \le |\eta| \le 2.4$ and the muons must have $E
\ge 2.1$ GeV in order to penetrate the absorber material.  The
identifier detectors also provide the necessary trigger information to
sample the muons from the high luminosity RHIC collisions.

The most substantial background in measuring $J/\psi$ and
$D$ mesons  are muons from pions and kaons that decay 
before they hit the brass nose-cone absorbers.  There is a competing
requirement in PHENIX in that one wants the absorber as close to the
interaction vertex as possible to reduce this decay contribution, but
they need to be far enough apart for there to be a good acceptance for
electrons, hadrons and photons in the two central arm spectrometers.
There are many benefits to the comprehensive design of PHENIX, but
there are definite drawbacks as well.  Of the two muon arms, the south
muon arm was completed for running in Run II at RHIC, and the north
arm will be complete for Run III.

\section{Physics Results}

In this chapter we present the physics results from the Run I data and 
a preview of expected results from Run II.

\subsection{Global Observables}

\begin{figure}[bth]
    \begin{center}
        \includegraphics[width=0.75\textwidth]{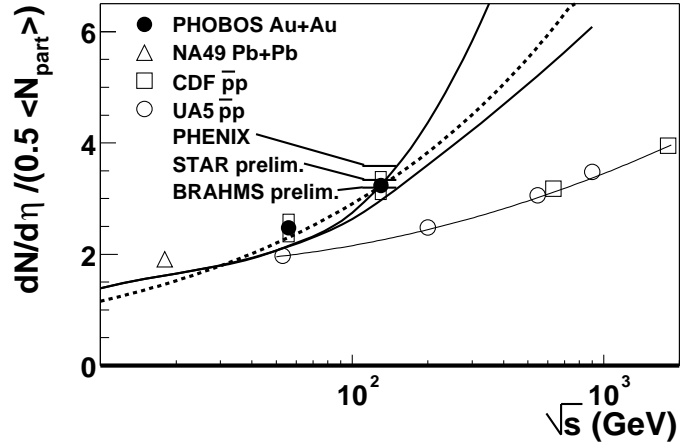}
    \end{center}
    \caption{Charged particle multiplicity measurement from all four RHIC
        experiments is shown
        for Au-Au collisions at \sqrtsNN\ = 130 GeV.  Also shown are
        data for $p+p$ and $p+\overline{p}$ collisions.  A model of heavy
        ion collisions HIJING is shown for comparison.}
    \label{fig:charge}
\end{figure}

The first result with a measurement from all four RHIC experiments is
the charged particle multiplicity
\cite{phobos-prl1,phenix-nch,brahms-qm,star-qm}. The four experiments'
results for central (small impact parameter) collisions are in
excellent agreement and are shown in Figure~\ref{fig:charge}.  The
multiplicity rises more sharply as a function of center-of-mass energy
in heavy ion collisions than in $p+p$ and $p+\overline{p}$ collisions,
which is attributed to the increased probability for hard parton
scattering in the thick nuclear target seen by each parton.

We expect the charge particle yield to increase for collisions of
larger nuclei.  However, at low x values, the high density of gluons
may in fact saturate due to gluon fusion processes.  The contribution
to the yield from hard processes should exhibit point-like scaling
(scaling with the number of binary collisions) and would thus scale as
$A^{4/3}$.  However, parton saturation depends upon the nuclear size
and would limit the growth of the number of produced partons as
$A^{1/3}$.  If present, this initial parton saturation would limit the
hard process contribution to the total charged particle multiplicity.

\begin{figure}[htb]
    \centerline{\includegraphics[width=0.75\textwidth] {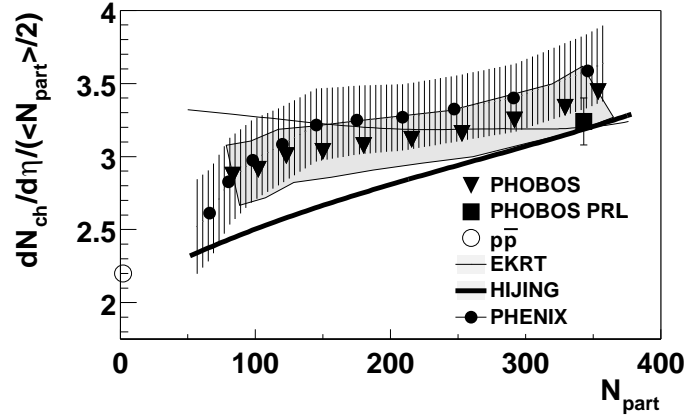}}
\caption{PHENIX and PHOBOS results for
    $dN_{ch}/d\eta|_{\eta=0} / {1 \over 2}N_{part}$ as a function of
    $N_{part}$.  The hashed and solid bands indicate the systematic
    errors for the two experimental results.  The data point for
    $p\overline{p}$ with two participants is shown for comparison.
    Also theoretical predictions from the HIJING and EKRT models are
    shown.}
\label{fig:npart}
\end{figure}

Only one nuclear species (Au) was accelerated in Run I at RHIC.  Thus,
rather that changing the mass number $A$ directly we control the
collision volume by varying the centrality or the number of
participating nucleons for Au-Au collisions. Shown in
Figure~\ref{fig:npart} are the published results from the PHENIX
\cite{phenix-nch} and PHOBOS \cite{phobos-nch2} experiments for the
number of charged particles per participant nucleon pair as a function
of the number of participating nucleons.  The number of participating
nucleons is determined in a slightly different manner by the different
experiments.  However, the general method is to calibrate the number
of spectator nucleons (= $2 \times A$ - participant nucleons) using a
measurement of spectator neutrons in a set of zero degree calorimeters
that are common to all experiments.  By correlating the number of
forward neutrons to the number of charged particle produced in the
large pseudo-rapidity region, the event geometry can be understood.

In Figure~\ref{fig:npart} one can also see theory comparisons that
indicate that a model including parton saturation (EKRT~\cite{ekrt})
fails to agree with the more peripheral data.  Results from the HIJING
model~\cite{hijing} are also shown which does not include parton
saturation and thus has a more continuous rise in the particle
multiplicity.  Since saturation phenomena are only likely to have
observable consequences for large collision volumes, it is not
possible with present systematics to rule out the saturation picture
for the most central collisions.

In order to better test the saturation picture lighter ion, smaller
$A$, collisions will be studied in Run II.  In addition, heavy flavor
(charm and bottom) and Drell-Yan production should be a sensitive
probe to the initial parton density.  Another proposal is that by
varying the collision energy and keeping the nuclear geometry the same
one can get a better handle on systematics and test scenarios
dependent on the coupling constant and the saturation scale.
The physics of parton saturation and color glass condensates is at the
forefront of theoretical development in the field.  Many recent 
developments were discussed in this workshop and will be described
in other contributions.


\begin{figure}
    \centerline{\includegraphics[width=0.77\textwidth] {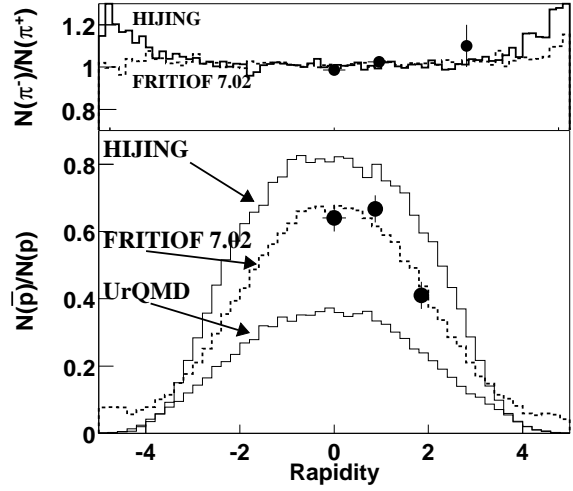}}
\caption{Plotted is the $\pi^{-}/\pi^{+}
    $ and $\overline{p}/p$ ratio as a function of rapidity from the
    BRAHMS experiment for central collisions.}
\label{fig:brahms_pbarp}
\end{figure}

In addition to the initial parton density, the energy
density is of great interest.  There are published results estimating
the initial thermalized energy density achieved in these collisions.
Bjorken originally derived a formula, shown in Eqn.~\ref{eqn:bj},
relating the measured transverse energy per unit rapidity to the
thermal energy density~\cite{bjorken}.
\begin{equation}
\epsilon_{B_{j}} = {{1} \over {\pi R^{2}}} {{1} \over {c\tau}} {{dE_\perp} \over {dy}}
\label{eqn:bj}
\end{equation}
It should be noted that there is a trivial factor of two error in the
original reference that is corrected here.  This formulation assumes a
boost invariant expanding cylinder of dense nuclear matter and a
thermalization time $\tau$.  There are two important assumptions in
this particular formulation.  The first is the boost invariant nature
of the collision.  There are recent preliminary measurements from STAR
and PHOBOS that indicate the distribution of particles is relatively
flat over $\pm 2$ units of pseudo-rapidity.  However, shown in
Figure~\ref{fig:brahms_pbarp} is the measured distribution of
$\overline{p}/p$ from the BRAHMS experiment~\cite{brahms-ratio}.  This
indicates the the system is already changing at $y \approx 2$,
though it is not clear that this is enough to invalidate the energy
density formulation.  The second question is what is the relevant
thermalization time $\tau$.

\begin{figure}
    \centerline{\includegraphics[width=0.75\textwidth] {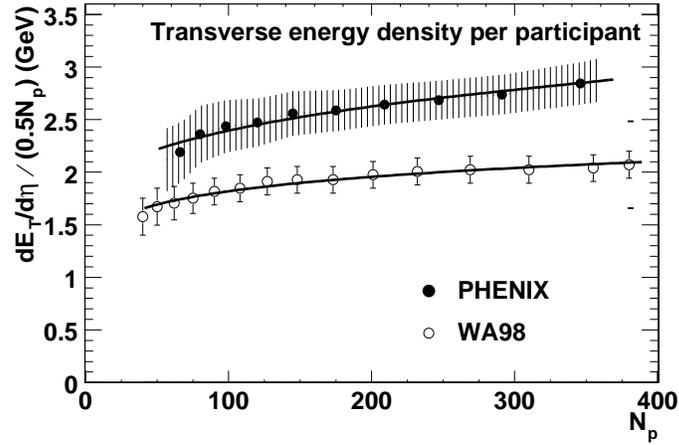}}
\caption{The PHENIX experiment result is shown for $dE_\perp/d\eta/(0.5N_{p})$
    at $\eta=0$ as a function of the number of participating nucleons.
    Also shown in the result from experiment WA98 at the lower energy
    CERN-SPS.}
\label{fig:phx-et}
\end{figure}

The PHENIX experiment has published \cite{phenix-et} the transverse
energy distribution for minimum bias Au-Au collisions.  For the 5\%
most central events, the extracted transverse energy
$<dE_\perp/d\eta>|_{\eta=0} = 503 \pm 2$ GeV.  Shown in
Figure~\ref{fig:phx-et} is $dE_\perp/d\eta/(0.5N_{p})$ versus the number
of participating nucleons.  One sees a similar increase in transverse
energy as was seen in the charged particle multiplicity yield.

The canonical thermalization time used in most calculations is
$\tau=1$ fm/$c$, that yields an energy density of $4.6$ GeV/fm$^{3}$,
which is is 60\% larger than measured at the CERN-SPS.  In addition,
it is believed that the density is substantially higher due to the
potentially much shorter thermalization time in the higher parton
density environment.  If one achieves gluon saturation the formation
time is of order 0.2 fm/$c$ and gives an estimated energy density of
$23.0~\mathrm{GeV/fm}^{3}$.  There are even estimates of over
$50~\mathrm{GeV/fm}^{3}$, but they assume a very large drop in the
final measured transverse energy due to work done in the longitudinal
expansion of the system.  All of these estimates are above the energy
density of order $0.6-1.8~\mathrm{GeV/fm}^{3}$ 
corresponding to the phase transition temperature $150-200$ MeV.

%
%

\subsection{Elliptic Flow}

The azimuthal anisotropy of the transverse momentum distribution for
non-central collisions is thought to be sensitive to the early
evolution of the system. The second Fourier coefficient of this
anisotropy, $v_{2}$, is called elliptic flow \cite{olliCorsica}. It is
an important observable since it is sensitive to the re-scattering of
the constituents in the created hot and dense matter. This
re-scattering converts the initial spatial anisotropy, due to the
almond shape of the overlap region of non-central collisions, into
momentum anisotropy. The spatial anisotropy is largest early in the
evolution of the collision, but as the system expands and becomes more
spherical, this driving force quenches itself. Therefore, the
magnitude of the observed elliptic flow reflects the extent of the
re-scattering at early time~\cite{sorge}.
\begin{figure}[htb]
    \begin{center}
        \includegraphics[width=0.78\textwidth]{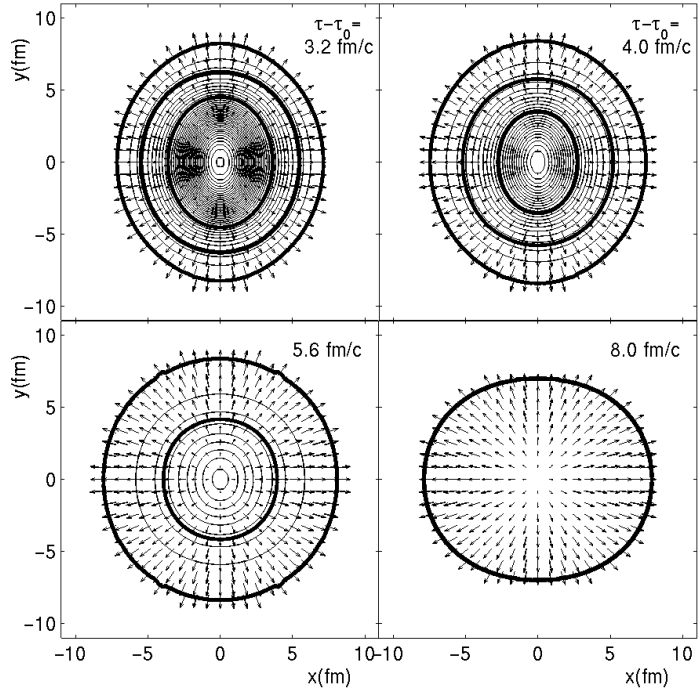}
\end{center}
\caption{Schematic view of a evolution of the transverse en\-
    ergy density profile (indicated by constant energy density con\-
    tours spaced by) and of the flow velocity field (indicated by
    arrows) for Pb+Pb collisions at impact pa\-rameter b=7.0 fm. The
    four panels show snapshots at times $\tau-\tau_0 =$ 3.2, 4.0, 5.6,
    and 8.0 fm/$c$. At these times the max\-imal energy densities in
    the center are 5.63, 3.62, 1.31 and 0.21 GeV/fm$^3$, respectively.
    The figure is taken from \protect\cite{uliflowPRC}.}
    \label{fig:uliflow}
\end{figure}
The time evolution of the transverse energy density profile is
schematically depicted in Figure~\ref{fig:uliflow} where the solid lines
represent surfaces of constant energy density.  The pressure in the
system is highest in direction of the reaction plane (largest energy
density gradient) which causes the elliptic anisotropy.

Elliptic flow in ultra-relativistic nuclear collisions was discussed
as early as 1992 \cite{olli92} and has been studied intensively in
recent years at AGS~\cite{e877flow2,e895},
SPS~\cite{na49prl,na49flow,wa98} and now at RHIC~\cite{starflow}
energies.  The studies at the top AGS energy and at SPS energies have
found that elliptic flow at these energies is in the plane defined by
the beam direction and the impact parameter, $v_2>0$, as expected from
most models.

The STAR detector is especially suited to study elliptic flow due to
its azimuthal symmetry, large coverage, and its capability of tracking
charged particles down to very low \pt. Even in peripheral events
there are sufficient tracks available to divide the event in two
'subevents' of which one is used to measure, or better estimate, the
event plane and the other to correlate the particles in it in order to
derive $v_2$. While BRAHMS is not able to study elliptic flow because
of its small acceptance, PHOBOS has some capabilities to measure $v_2$
but only integrated over all \pt.  Because of its restricted azimuthal
coverage the PHENIX collaboration follows a different approach to
reconstruct $v_2$ by studying the $\Delta \phi$ correlation between
particles thus circumventing the event plane determination.  However,
work is still in progress and in the following we concentrate on the
more direct event plane method used in STAR.

The flow analysis method involves the calculation of the
event plane angle, which is an experimental estimator of the real
reaction-plane angle.  The second harmonic event plane angle,
$\Psi_2$, is calculated for two sub-events, which are independent
subsets of all tracks in each event.
\begin{figure}[hbt]
    \begin{center}
        \includegraphics[width=0.8\textwidth]{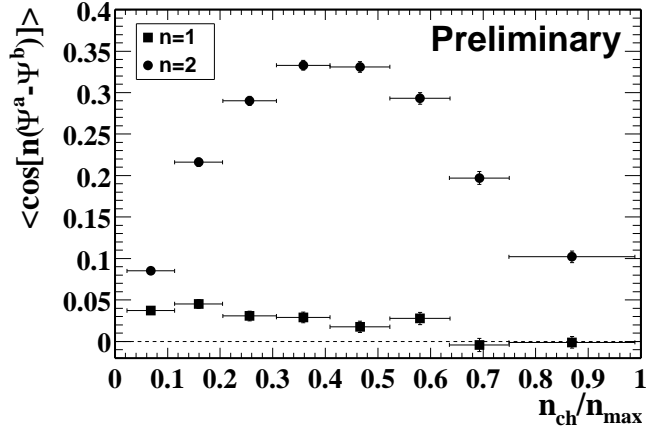}
\end{center}
\caption{The correlation between the event plane 
        angles determined for two independent sub-events. The
        correlation is calculated for the first harmonic (n=1) and the
        second harmonic (n=2).}
    \label{resolution}
\end{figure}
Figure~\ref{resolution} shows the results for the correlation between
the sub-events for the first and second harmonic as a function of
centrality~\cite{starflow}. The peaked shape of the centrality
dependence of $\langle\cos[2(\Psi_a - \Psi_b)]\rangle$ is a signature
of anisotropic flow. However, the correlation between the sub-events
may not be due entirely to anisotropic flow. To estimate the magnitude
of non-flow effects one can use the sub-events in three different
ways:

\begin{enumerate}
\item Assigning particles with pseudo-rapidity $\eta < 0+\epsilon$ to
    one sub-event and particles with $\eta > 0+\epsilon$ to the other.
    Short range correlations, such as Bose-Einstein or Coulomb, are to
    a large extent eliminated by the $2 \epsilon$ ``gap'' between the two
    sub-events.
\item Dividing randomly all particles into two sub-events, sensitive
    to all non-flow effects.
    
\item Assigning positive particles to one sub-event and negative
    particles to the other, allowing an estimation of the contribution
    from resonance decays.
\end{enumerate}

Studies have shown that the results from all three methods are for the
central and mid-peripheral events very similar. For the most
peripheral events the results vary among the methods by about 0.005.
Not all non-flow contributions might be known and the effects of
others, such as jets, are difficult to estimate because of their
long-range correlation.  In order to estimate the systematic
uncertainty due to the effects of jets, one can assume that jets
contribute at the same level to both the first and second order
correlations.  Taking the maximum observed positive first order
correlation, as being completely due to non-flow will reduce the
calculated $v_2$ values.
\begin{figure}[hbt]
    \begin{center}
        \includegraphics[width=0.8\textwidth]{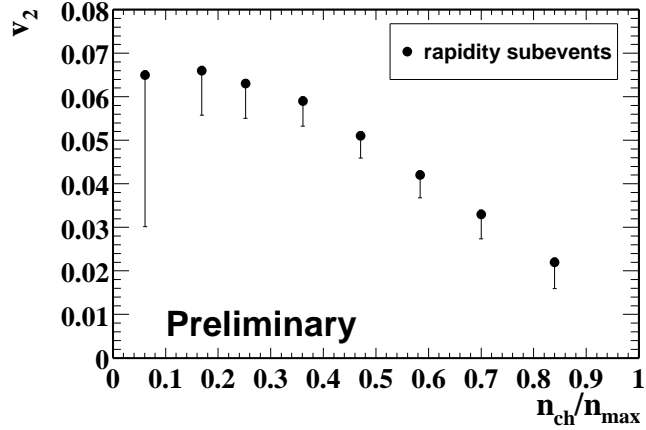}
\end{center}
\caption{The integrated elliptic flow signal, $v_2$, with the
        estimated systematic uncertainties as measured by STAR.}
    \label{errors}
\end{figure}

\begin{figure}[hbt]
    \begin{center}
    \includegraphics[width=0.8\textwidth]{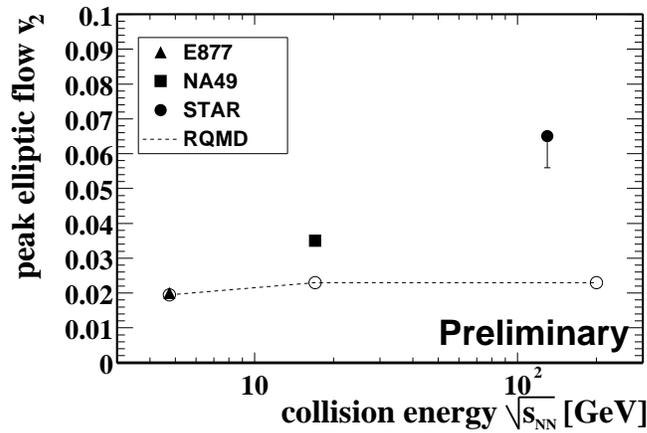}
\end{center}
    \caption{Excitation function of $v_2$ from top AGS to RHIC energies.}
    \label{exitation}
\end{figure}

Figure~\ref{errors} shows the final $v_2$ integrated over all \pt\ as
a function of centrality. The statistical uncertainties are smaller
than the markers and the uncertainties shown are the systematic
uncertainties due to this estimated non-flow effect.

Figure~\ref{exitation} shows the maximum $v_2$ value as a function of
collision energy. It rises monotonically from about 0.02 at the top
AGS energy~\cite{e877flow2}, 0.035 at the SPS~\cite{na49flow} to about
0.06 at RHIC energies~\cite{starflow}. This increasing magnitude of
the integrated elliptic flow indicates that the degree of
thermalization, which is associated with the amount of re-scattering,
is higher at the higher beam energies. However, interpretation of the
excitation function has to be done with care. The $v_2$ values used
here are the maximum values as a function of centrality for each
energy.  The centrality where $v_2$ peaks can change as a function of
beam energy, indicating different physics~\cite{physeflow}.
 
\begin{figure}[hbt]
    \begin{center}
        \includegraphics[width=0.8\textwidth]{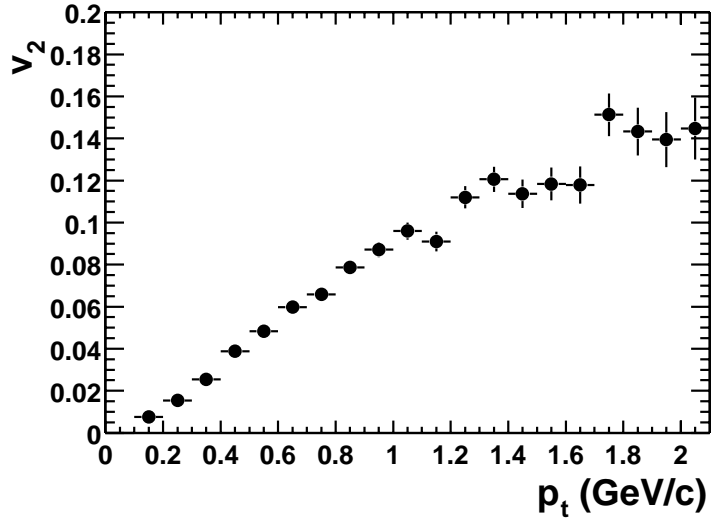}
\end{center}
\caption{$v_2$ as a function of \pt, as measured by STAR
    in \sqrtsNN\ = 130  GeV Au-Au collisions.}
    \label{flowpt}
\end{figure}

The differential anisotropic flow is a function of $\eta$ and \pt.
Figure~\ref{flowpt} shows $v_2$ for charged particles as a function of
\pt\ for a minimum bias event sample.  Mathematically, the $v_2$ value at
\pt=0, as well as its first derivative, must be zero, but it is
interesting that $v_2$ appears to rise almost linearly with \pt\ 
starting from relatively low values of \pt\.  This is consistent with a
stronger ``in-plane'' hydrodynamic expansion of the system than the
average radial expansion.

Comparing to estimates based on transport cascade models, one finds
that elliptic flow is under predicted by a factor of more than 2.
Hydrodynamic calculations~\cite{uliflowPRC} for RHIC energies
over predict elliptic flow by about 20-50\%. This is just the reverse
of the situation at the SPS where cascade models gave a reasonable
description of the data and hydrodynamic calculations were more than a
factor of two too high. Also in contrast to lower collision energies,
the observed shape of the centrality dependence of the elliptic flow
is similar to hydrodynamic calculations and thus consistent with
significant thermalization which is one of the most striking results
from the initial round of RHIC results.

%
%
\subsection{Two-Particle Interferometry (HBT)}

The study of small relative momentum correlations, a technique also
known as HBT~\cite{hbt} interferometry, is one of the most powerful
tools to study complicated space-time dynamics of heavy ion collisions
\cite{hbtreview}.  It provides crucial information which helps to
improve our understanding of the reaction mechanisms and to constrain
theoretical models of the heavy ion collisions.  Interpretation of the
extracted HBT parameters in terms of source sizes and lifetime is more
or less straightforward for the case of chaotic static sources. In the
case of expanding sources with strong space-momentum correlations (due
to flow, etc.) the situation is more difficult, but the concept of
length of homogeneity~\cite{makhlin_sinyukov} provides a useful
framework for the interpretation of data.

The dependence of the pion-emitting source parameters on the
transverse momentum of the particle pairs ($K_T$) and on centrality
can in principle be measured by all RHIC experiments with high
statistics. For more detailed analysis as for example event-by-event
HBT, HBT radii versus reaction plane, and the correlation of HBT
results with other observables can only be performed by STAR due to
its large acceptance and azimuthal coverage. These studies, however,
are still in progress and it is by far too early to discuss them here.
In the following we show results from the STAR experiment
\cite{star_hbt_paper} that performed a multi-dimensional analysis
using the standard Pratt-Bertsch decomposition \cite{Pratt} into
outward, sideward, and longitudinal momentum differences and radius
parameters.  The data are analyzed in the longitudinally co-moving
source frame, in which the total longitudinal momentum of the pair
(collinear with the colliding beams) is zero.

\begin{figure}[p]
    \begin{center}
        \includegraphics[height=0.8\textheight]{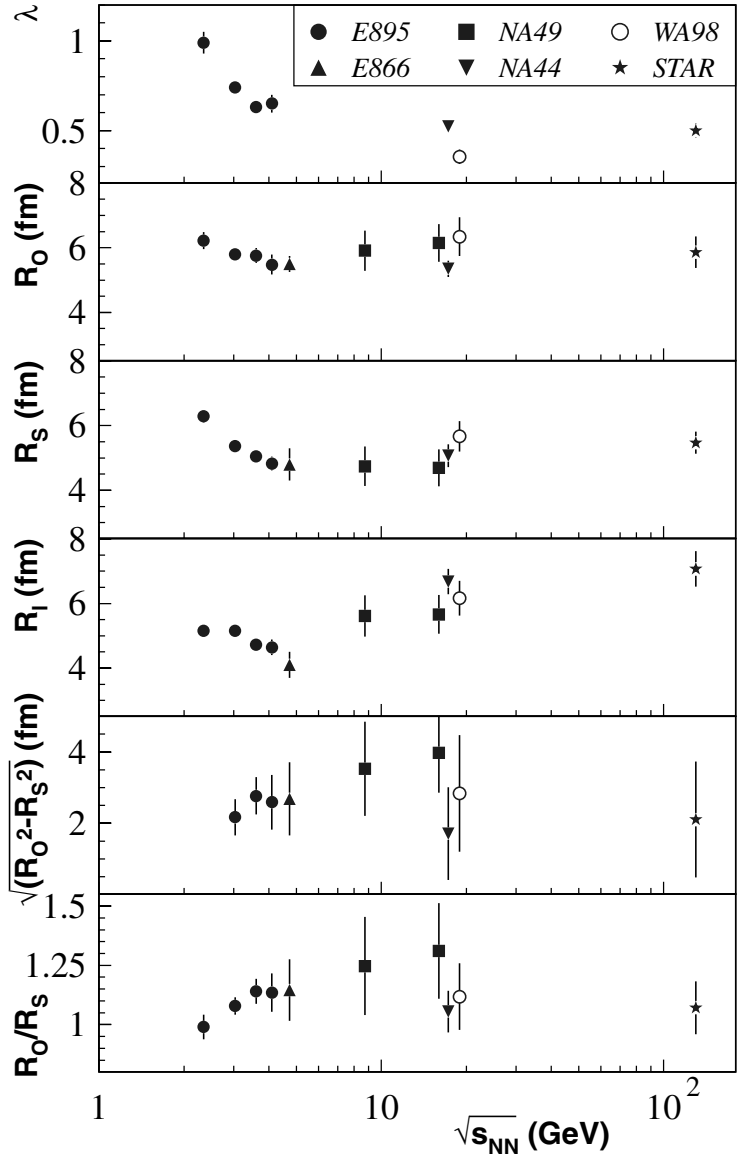}
\end{center}
\caption{Compilation of results on two-particle correlation (HBT) parameters
    from measurements using central collisions of Au + Au at the
    BNL-AGS, Pb + Pb at the CERN-SPS and Au + Au data from the STAR
    experiment at RHIC. Plotted are the coherence parameter $\lambda$,
    R$_{out}$, R$_{side}$, and R$_{longitudinal}$.}
\label{fig:HBT}
\end{figure}

As expected, larger sizes of the pion-emitting source are found for
the more central (\textit{i.e.} decreasing impact parameter) events,
which in turn have higher pion multiplicities.  This source size is
observed to decrease with increasing transverse momentum of the pion
pair.  This dependence is similar to what has been observed at lower
energies and is understood to be an effect of collective transverse
flow. Shown in Figure~\ref{fig:HBT} is the coherence parameter $\lambda$
and the radius parameters R$_{out}$, R$_{side}$, and R$_{long}$
obtained in the analysis.  Also shown are values of these parameters
extracted from similar analyses at lower energies.  All analyses are
for low transverse momentum ($\sim$ 170 MeV/$c$) negative pion pairs at
mid-rapidity for central collisions of Au + Au or Pb + Pb.  From
Figure~\ref{fig:HBT} the values of $\lambda$, R$_{out}$, R$_{side}$, and
R$_{long}$ extend smoothly from the dependence at lower energies and
do not reflect significant changes in the source from those observed
at the CERN SPS energy.  One of the biggest surprises is that the
anomalously large source sizes or source lifetimes predicted for a
long-lived mixed phase \cite{Rischke} have not been observed in this
study.  Preliminary results of the HBT analysis by the PHENIX
Collaboration ~\cite{johnson} agree with the STAR results within error
bars.

One of the big puzzles, however, is the magnitude and the tranverse
momentum ($K_T$) dependence of the ratio of $R_{out}$/$R_{side}$ which
contradicts \emph{all} model predictions~\cite{Rischke,soff_bass_dumitru}.
These model calculations predict the ratio to be greater than unity
due to system lifetime effects which cause $R_{out}$ to be larger than
$R_{side}$.  They also predict that the ratio increases with $K_T$.
Such an increase seems to be a generic feature of the models based on
the Bjorken-type, boost-invariant expansion scenario. Hence, it was
surprising to see that the experimentally observed ratio is less than
unity and is decreasing as a function of $K_T$. Currently, it is far
from clear what kind of scenario can lead to such a puzzling $K_T$
dependence.

\subsection{Hard Process Probes of the Plasma}

\begin{figure}[bht]
    \centerline{\includegraphics[width=0.95\textwidth]
        {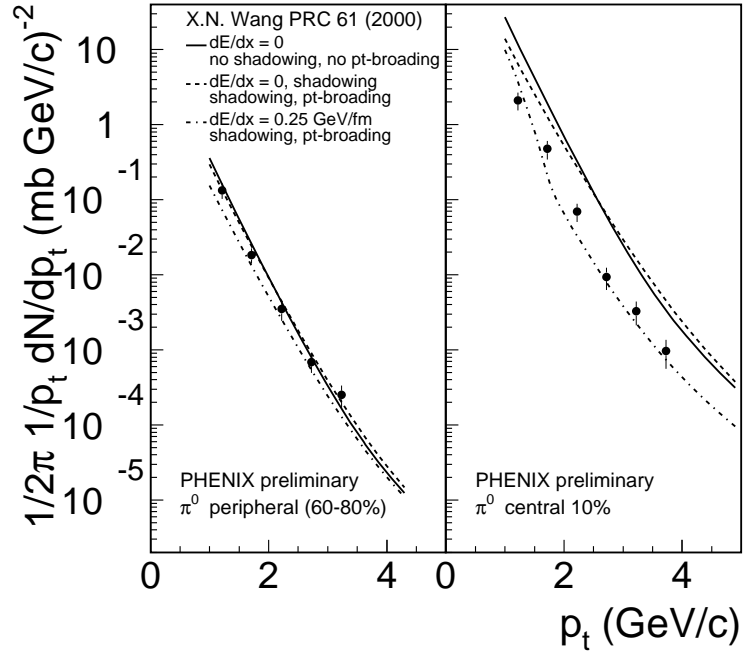}}
  \caption{Preliminary PHENIX invariant multiplicity of identified
      $\pi^{0}$ as a function
      of transverse momentum are shown for peripheral and central
      collisions.  Comparison with theoretical calculations with and
      without parton energy loss are also shown.}
   \label{fig:phx-pizero}
\end{figure}

Jet processes and their associated hadronic fragmentation provide one
of the most exciting probes of the color deconfined plasma.  The
PHENIX experiment has measured the distribution of identified
$\pi^{0}$ for both central and peripheral Au-Au collisions as shown in
Figure~\ref{fig:phx-pizero}~\cite{phx-qm}.  The peripheral results
appear to be in good agreement within systematic errors of an
extrapolation from $pp$ collisions scaled up by the number of binary
collisions expected in this centrality class.  However, the central
collision results show a significant suppression in the $\pi^{0}$
yield relative to this point like scaling expected for large momentum
transfer parton-parton interactions.  If the created fireball in RHIC
collisions is transparent to quark jets, then we expect the yield of
high \pt\ hadrons to obey point-like scaling and equal the $pp$ (or
equivalently $p\overline{p}$) distribution scaled up by the number of
binary $NN$ collisions, or equivalently by the nuclear thickness
function $T_{AA}$.  This is not what is observed.  A more
sophisticated calculation \cite{wang-prc} yields the same qualitative
conclusion.

\begin{figure}[bt]
    \centerline{\includegraphics[width=0.82\textwidth] {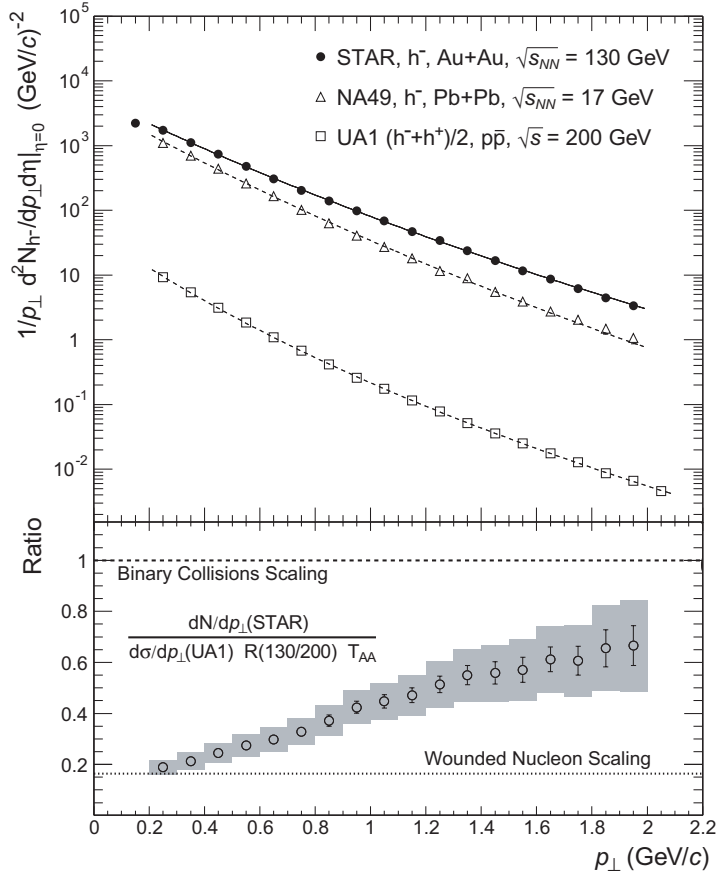}}
  \caption{Invariant multiplicity of charged hadrons as a
      function of transverse momentum (top).
      Ratio of unidentified charge hadrons per calculated binary
      collision from Au + Au central collisions to those from p +
      p($\overline{p}$) collisions extrapolated to \sqrtsNN\,=130 GeV
      as a function of transverse momentum (bottom).}
  \label{fig:star_pt}
\end{figure}

The STAR experiment has recently submitted for
publication~\cite{star-highpt} the \pt\ spectra for unidentified
negatively charged hadrons in central Au + Au collisions as shown in
Figure~\ref{fig:star_pt}.  Also shown are the equivalent spectra from
experiment NA49 at the CERN-SPS at $\sqrt{s}=17$ GeV and from UA1 in
$p\overline{p}$ at \sqrtsNN\, = 200 GeV.  The STAR spectra is then
divided by the spectra from $p\overline{p}$ scaled by the number of
binary collisions, and the result is shown in the lower panel of
Fig~\ref{fig:star_pt}.  At low transverse momentum the particle
production is dominated by soft interactions which scale with the
number of wounded nucleons as indicated by the line at 0.2.  The rise
from 0.2 as a function of \pt\ certainly has a large contribution
from hydrodynamic flow that will push particles to higher transverse
momentum in central Au-Au collisions.

\begin{figure}
    \centerline{\includegraphics[width=0.75\textwidth] {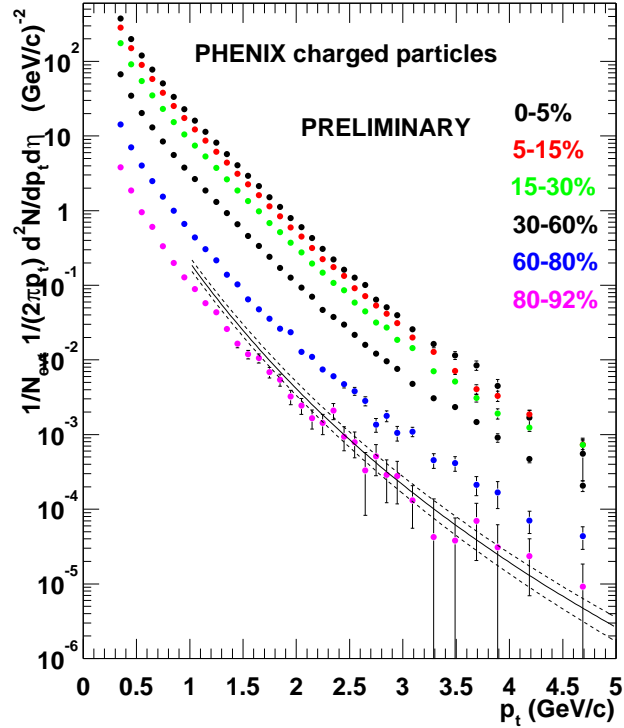}}
  \caption{PHENIX preliminary results for unidentified charged hadron
      invariant multiplicity as a function of transverse momentum.}
  \label{fig:phx-spectra}
\end{figure}

\begin{figure}
    \centerline{\includegraphics[width=0.95\textwidth]
        {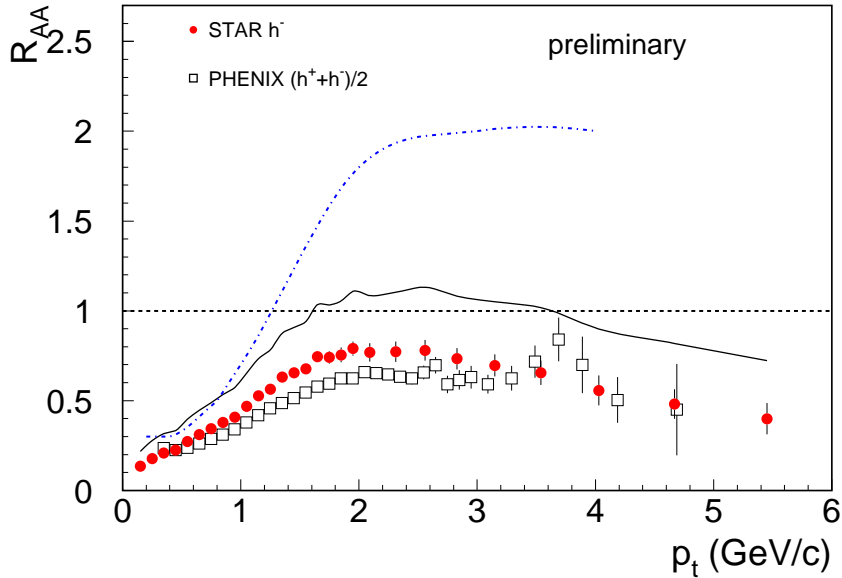}}
  \caption{Ratio of unidentified charge hadrons per calculated binary
      collision from Au + Au central collisions to those from p +
      p($\overline{p}$) collisions (extrapolated to \sqrtsNN\, = 130
      GeV) as a function of transverse momentum (GeV).  The solid line
      is the systematic error band on the ratio.  The dashed line is
      the average result from experiments at the lower energy
      CERN-SPS.}
  \label{fig:highpt_ratio}
\end{figure}

The PHENIX experiment has shown preliminary results extending out
further in transverse momentum.  Preliminary results for six
centrality classes are shown from PHENIX in Fig~\ref{fig:phx-spectra}.
STAR also has preliminary results for central collisions extending out
to $\pt\, > 5$ GeV that are in reasonable agreement with the PHENIX
results.  If one takes the ratio of the central spectra to the
unidentified spectra in $pp$ collisions scaled up by the number of
binary collisions one gets a ratio $R_{AA}$ as shown in
Fig~\ref{fig:highpt_ratio}.  It needs to be noted that there is no
$pp$ data at \sqrtsNN\, = 130 GeV and thus an extrapolation to that
energy is done to calculate $R_{AA}$.  This extrapolation is included
in the systematic error band, and should be reduced when both
experiments measure the spectra in $pp$ in Run II.

There are many important physics points to understand in these
results.  The ratio appears to stay below one, although that is a
marginal conclusion with the present systematic errors.  However, this
is certainly in qualitative agreement with a parton energy loss
scenario, as also seen in the observed suppression in the PHENIX
$\pi^{0}$ spectra.  In contrast, the CERN-SPS results show an
enhancement that has been attributed to the Cronin effect, or initial
state parton scattering that gives a $k_T$ kick to the final
transverse momentum distribution.  This expected enhancement makes the
suppression seen at RHIC all the more striking.

There are a number of open questions that must be considered before
drawing any conclusions.  The most basic is that these \pt\ values
are low relative to where one might have confidence in the
applicability of perturbative QCD calculations.  In addition, the
separation between soft and hard scale physics is blurred in this
\pt\ range, and in fact the CERN-SPS ratio $R_{AA}$ has also been
explained in terms of hydrodynamic boosting of the soft physics to
higher \pt.  The preliminary results from PHENIX on the ratios of
$\pi/K/p(\overline{p})$ in the middle of this \pt\ range look more
like soft physics than a parton fragmentation function in vacuum.  One
additional point of concern is that these models of energy loss assume
that the parton exits the collision region before finally fragmenting
into a jet of forward hadrons.  Thus the final hadronization takes
place in vacuum.  In the \pt\ range of these early measurements, that
conclusion is not so clear.  The parton is traveling through the
medium with various $k_T$ scatters, and if it hadronizes inside a bath
of other particles, the leading hadrons may be slowed down by
inelastic collisions with co-moving pions.  Lastly, the point-like
scaling is known to be violated due to the nuclear shadowing of parton
distribution functions.  These nuclear modifications are known to
reduce the pdf for quarks of order 20\% for $x\approx10^{-2}$;
however, the shadowing for gluons is not currently measured.  The
calculations of \cite{wang-prc} have included modeling of this
shadowing, but must be viewed with caution at this time.  These points
need further theoretical investigation.  In addition, as will be
discussed in the next section, many of these concerns are reduced when
the measurements extend to much higher transverse momentum.

\section{Future Measurements}

In the following chapter, we discuss a few select topics within our
areas interest that have exciting results expected in the near future.

\subsection{Charm and Bottom}

The measurement of open charm and bottom in relativistic heavy ion
collions is both an extreme experimental challenge and rich with
physics information.  First, the measurement of quarkonium states such
as the $J/\psi$ require a comparison measure of the original
$c\overline{c}$ production to determine the effect of color screening.
Also, the total charm production is sensitive to the initial gluon
density in the incoming nuclei and is thus sensitive to any shadowing
of the gluon distribution function and may even comment on the
possible color glass condensate postulated to describe the phase space
saturated gluon distributions in the highly Lorentz contracted nuclei.
Lastly, recent predictions of charm quark energy loss in tranversing a
hot partonic medium have generated much interest.  Now for the
difficult part.  The best was to measure charm via D mesons is either
via direct reconstruction from its $\pi + K$ decay mode or via the
semi-leptonic decay $K + e + \nu_{e}$.  The combinatoric background in
the purely hadronic channel are close to overwhelming and the
semi-leptonic decay cannot be completely reconstructed.  In particle
physics experiments measuring the decay products with a few micron
displaced vertex from the collision vertex allows for a dramatic
reduction in the combinatoric background.  However, the level of
silicon detector technology was not advanced enough for this very high
multiplicity environment at the time of the RHIC detector designs.
This is now being discussed as a possible upgrade to the experiments.

\begin{figure}
    \centerline{\includegraphics[width=0.85\textwidth] {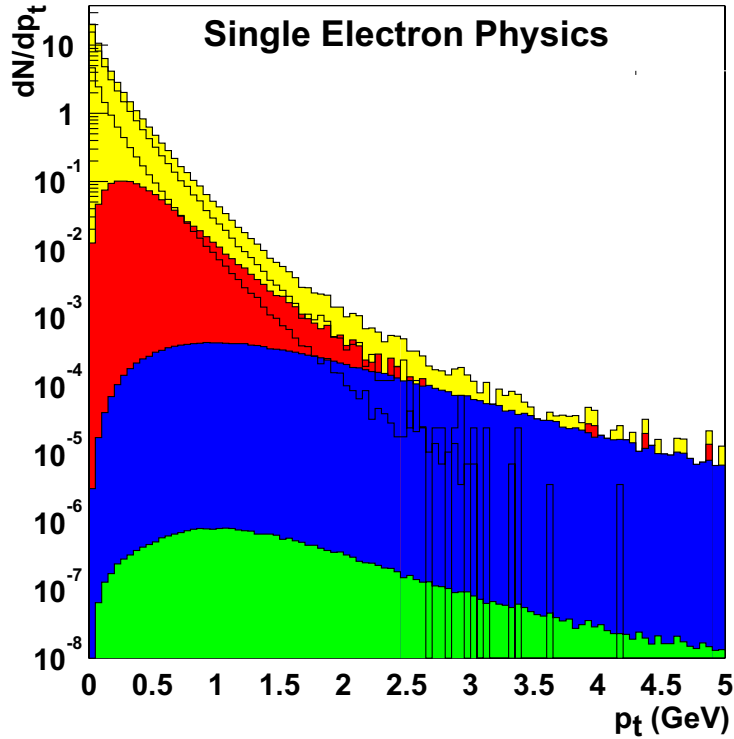}}
  \caption{Simulation of the transverse momentum spectrum for mid-rapidity
      single electrons and positrons.  The top curve is the sum of all
      contributions.  the two lower curves are from Dalitz decays and
      photon conversions.  The subsequent three grey shaded areas are
      from charm D mesons, beauty B mesons, and Drell-Yan.}
  \label{fig:singlee}
\end{figure}

One promising way of determining the charm production is through the
measurement of single electrons.  Shown in Figure~\ref{fig:singlee} is
a simulation of single electrons as a function of transverse momentum.
The top curve in the sum of all contributions.  The next two curves
that dominate the spectra at low $\pt\, < 1.0$ GeV and from the Dalitz
decays of pions and $\eta$ and from photon conversions.  The next two
contributions are from charm D meson and beauty B meson decays.  Charm
yields $\approx 50\%$ of the counts at \pt\, =  GeV, and beauty
yields $> 50\%$ of the counts above $\pt\, > 3.5$ GeV.  The lowest
curve is the contribution from Drell-Yan which never has a major
contribution to the single electrons.

\begin{figure}
    \centerline{\includegraphics[width=0.85\textwidth] {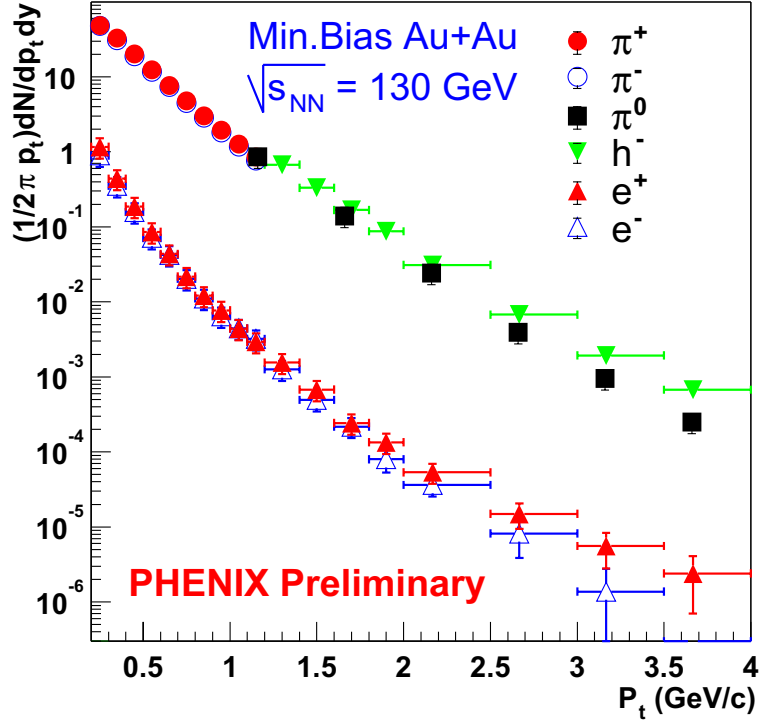}}
  \caption{PHENIX preliminary results for unidentified charged hadron
      invariant multiplicity as a function of transverse momentum.}
  \label{fig:elec}
\end{figure}

PHENIX has made a preliminary measurement of the single electron
transverse momentum spectra from the limited statistics in Run I as
shown in Figure~\ref{fig:elec}.  The analysis of these results is
proceeding and implications on charm production are forthcoming.

Additional handles on heavy flavor production can be had with the
measurement of correlated leptons.  For example, electron-muon pairs
at large relative momentum ($Q^{2}$ or $M_{inv}$) have a substantial
contribution from $c\overline{c}$ and $b\overline{b}$ pairs.  For
example, the $c \overline{c}$ can fragment into $D \overline{D}$
followed by the decays $D \longrightarrow K + e + \nu_{e}$ and
$\overline{D} \longrightarrow K + \mu + \nu_{\mu}$.  Although the
initial $Q^{2}$ of the $c\overline{c}$ pair is significantly modified
when measured as a $Q^{2}$ of the $e \mu$ pair, there is enough
information to attempt to extract a total charm cross section and
maybe something about the initial $Q^{2}$ of the $c\overline{c}$.
This measurement has the advantage over $e^{+}e^{-}$ and
$\mu^{+}\mu^{-}$ pairs in that $e\mu$ pairs are free from Drell-Yan
and thermal contributions.

One of us (J.N.) has used PYTHIA 6.0 to estimate the rate of $e\mu$
pairs into the PHENIX acceptance (Central arms for the electron and
South muon arm only for the $\mu$).  PYTHIA has been run with a charm
quark mass of 1.5 GeV/$c^{2}$ and $<\langle k_{T} \rangle$ = 1.5 \gevc.

\begin{figure}[thb]
    \centerline{\includegraphics[width=0.80\textwidth]{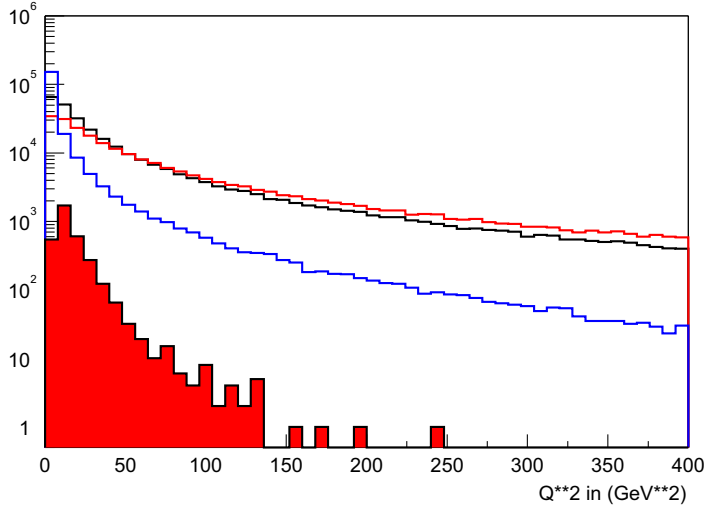}}
\caption{$Q^{2}$ in GeV$^{2}$ for $c\overline{c}$ (black line),
    $D\overline{D}$ (red line), 
    $e\mu$ (blue line), and $e\mu$ accepted by PHENIX (red fill).  The
    vertical scale is arbitrary between the different curves.}
\label{fig:q2compare}
\end{figure}
\begin{figure}[h]
    \centerline{\includegraphics[width=0.82\textwidth]{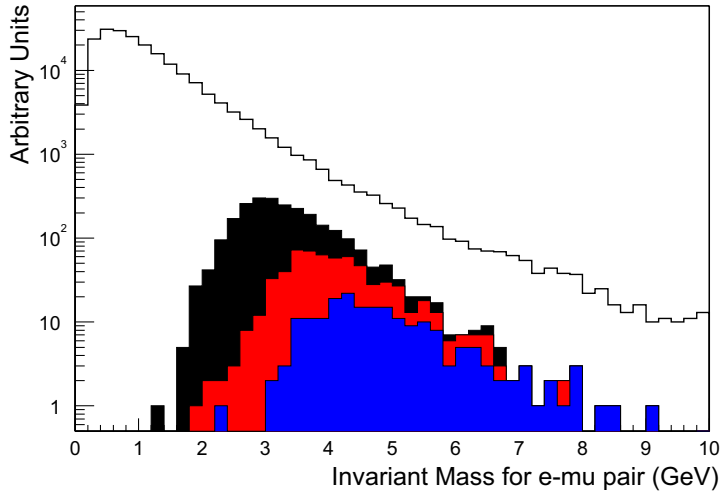}}
\caption{$e\mu$ Invariant Mass Distribution in GeV.  
    Shown are all $e\mu$ pairs (black line), $e\mu$ pairs with the
    muon penetrating to gap 5 and $E(\mathrm{electron})>1.0$ GeV (black fill),
    with $E(\mathrm{electron})>1.5$ GeV (red fill), $E(\mathrm{electron})>2.0$
    GeV (blue fill).}
\label{fig:invmass}
\end{figure}
We show in Figure~\ref{fig:q2compare} the distribution of $Q^{2}$ for
all $c\overline{c}$, $D\overline{D}$, $e\mu$ and $e\mu$ pairs accepted
by PHENIX.  One can see that the charm mesons carry most of the
information from the $c\overline{c}$ pair.  For the $e\mu$ pair the
correlation with the $c\overline{c}$ pair $Q^{2}$ is substantially
washed out and of much lower slope since the kaon in the D decay takes
away a large fraction of the original charm momentum that is not
measured.  It should be noted that the modeling of the fragmentation
of the charm quark will be a source of systematic error, whereas the
blurring of the $Q^{2}$ from the $e\mu$ decay kinematics can be
modeled exactly.

We show the invariant mass distribution of the $e\mu$ pairs into the
PHENIX acceptance in Figure~\ref{fig:invmass}.  Also shows are three
mass distribution with electron energy cuts of 1.0, 1.5, and 2.0 GeV.
The cut on the electron energy is strongly correlated with the
invariant mass selection on the $e\mu$ pair.  Since lower masses (in
particular $m < 4.0$ GeV) are thought to have large background
contamination, the loss of these pairs with higher electron energy
threshold are not as worrisome as they might otherwise be.  It may be
necessary to use the electron energy in order to selectively trigger
on these events.

\begin{figure}[htb]
    \centerline{\includegraphics[width=0.90\textwidth]{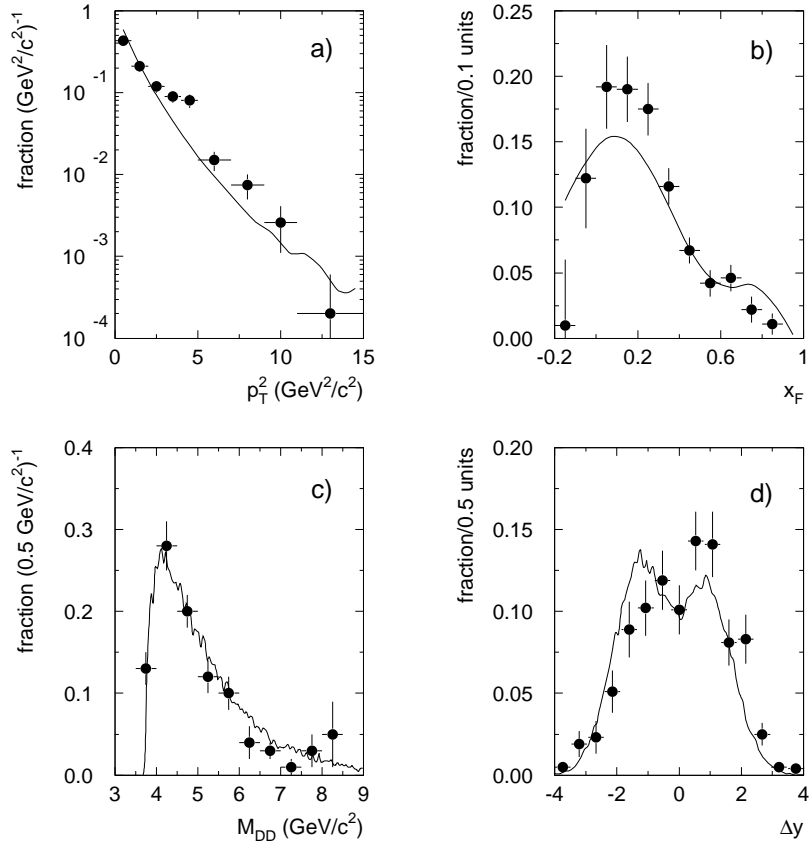}}
\caption{$D\overline{D}$ data from $\pi^{-}-Cu$ collisions at $\sqrt{s}=26$ GeV.
    In particular, panel d) shows the $D\overline{D}$ rapidity
    difference $\Delta y = y_{D} - y_{\overline{D}}$.}
\label{fig:pythia_data_deltay}
\end{figure}

Measuring charm by this method requires an accurate model of the
$c\overline{c}$ distribution in $Q^{2}$, and the corresponding $\Delta
y$ (rapidity gap) and $\Delta \pt$.  In Figure
~\ref{fig:pythia_data_deltay} PYTHIA is compared to some of the only
data on the rapidity gap between $D$ and $\overline{D}$
mesons \cite{drees}. The agreement is not bad, but it the comparison
is not a great confidence builder in the ability to model and then
test by data checking this input.  More experimental data are needed
and theoretical work to model charm production.

In order to draw a full picture of charm production, multiple
measurements must be done in the single lepton (electrons and muons)
and correlated leptons (electron pairs, muon pairs, and electron-muon
pairs).  It would be extremely useful for theorists interested in
total charm or beauty production, and also high \pt\ energy loss of
heavy flavor partons, to make some predictions for the leptonic
signatures that will be measured in the next year.  Future upgrades to
the detectors for tagging displaced vertices from D meson decays are
probably more than five years away.

\subsection{Quarkonia (PHENIX)}

\begin{figure}[htb]
    \centerline{\includegraphics[width=0.80\textwidth]{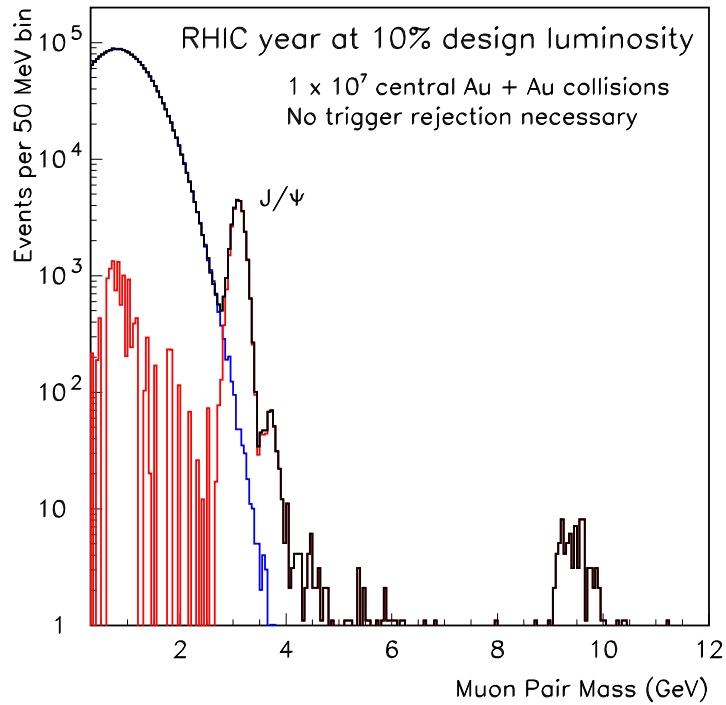}}
\caption{Simulation study of dimuons in the PHENIX muon spectrometers.}
\label{fig:jpsi}
\end{figure}

There are no early measurement results on quarkonia $J/\psi$ states
from Run I due to the low luminosity and short running period.  The
PHENIX experiment will make a measurement of $J/\psi$ and other states
in both the muon and electron channels in Run II.  PHENIX has a large
acceptance in $x_{F}$ and \pt\ that we be crucial to constraint
models of coloring screening absorbtion and test theories with
re-coalescence at the hadronization phase.  Shown in
Figure~\ref{fig:jpsi} is a simulation of the type of measurement that
could be made in the PHENIX muons arms with 37 weeks at 10\% of design
luminosity, or equivalently in less than four weeks with the
luminosity averaging the the RHIC design specification.  The first
measurements are being made now in Run II, and high statistics should
be available in Run III.  

\subsection{Quarkonia (STAR)}

The STAR detector system is unique among the RHIC experiments in its
capabilities to simultaneously measure many experimental observables
on a event-by-event basis such as energy-density, entropy,
baryochemical potential, strangeness content, temperature, and flow.
The measurement of \jpsi\ production as a function of these quantities
allows the study of the suppression mechanism in great detail.  This
advantage becomes immediately apparent when it comes to the study of
the \textit{onset} of the anomalous suppression which reflects the
point where the system reaches critical conditions and, at least
partially, undergoes a phase transition. Thus, the correlation of the
$c\bar{c}$ break-up with the many single-event variables provides a
new promising analysis tool.

The golden \jpsi\ decay mode for STAR is $\jpsi \rightarrow \ep$.
Prima facie this poses a problem since the STAR detector has been
designed to focus primarily on hadronic observables over a large
phase-space and thus lacks two essential features of a dedicated
lepton experiment: hadron-blind detectors and fast event recording
rates. The two essential components which help to overcome these
shortcomings are \textit{(i)} the EMC barrel which allows to suppress
hadrons to a level sufficient to achieve signal-to-background ratios
around 1:3 or better, and \textit{(ii)} a fast level-3 trigger which
is designed to efficiently trigger on electron-positron pairs with a
given invariant mass at rates in the order of 100 Hz, thus improving
STARs bandwidth for recording \jpsi\ decays by almost two orders of
magnitude.

\begin{figure}[hbt]
    \begin{center}
        \includegraphics[width=\textwidth]{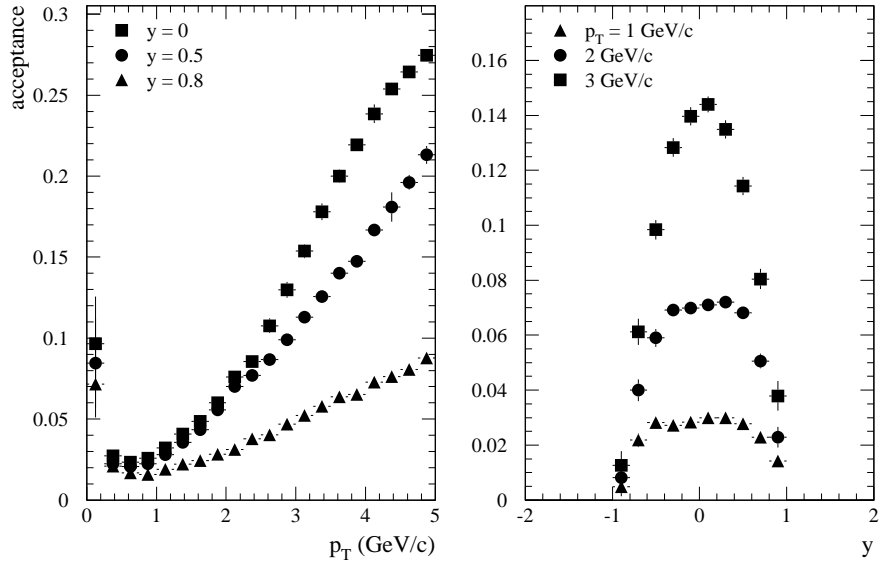}
    \end{center}
        \caption{Geometrical acceptance for $\jpsi \rightarrow \ep$
            in the STAR experiment. The left plot shows the acceptance as a
            function of \pt\ for various rapidities $y$, right one as
            a function of rapidity for various \pt\ slices.}
        \label{fig:acceptance}
\end{figure}

The geometric acceptance for the decay channel $\jpsi \rightarrow \ep$
in STAR is shown in Fig.~\ref{fig:acceptance}. A \jpsi\ is accepted if
both electrons carry momenta $\mathrm{p} > 1.5\ \gevc$ and fall into
the EMC acceptance.  Both requirements imply that the electron tracks
cross all layers of inner tracking detectors (SVT+SSD) and all
TPC padrows. This ensures maximum momentum and dE/dx resolution and
therefore maximum additional electron identification from detectors
other than the EMC.  As is depicted in the left plot, the full
coverage of the EMC $|\eta| < 1$ ensures relatively high efficiencies
($\sim 8\%$) at very low \pt.  At these low values the two leptons run
essentially back-to-back, a decay topology that requires symmetric
coverage around $y = 0$.  This region of phasespace is of great
interest since here the \jpsi\ remains longest in the hot dense medium
and its breakup probability is maximal. Up to $\pt\ = 1\ \gevc$ the
acceptance then drops significantly since, still at large opening
angles, it becomes more likely to loose one electron because it either
carries too low momentum and/or falls outside our acceptance. At
larger \pt\ the acceptance raises dramatically due to the decreasing
opening angle of the pair an the higher average momenta of the
electrons.  However, the \jpsi\ cross-section drops exponentially
towards larger \pt\ resulting in little net benefit in terms of total
yields.  In this region the acceptance scales approximately
\textit{linear} with the EMC coverage while at lower \pt\ it scales
almost \textit{quadratically}.

The dominant background source is the 'combinatorial background'
due to pions misidentified as
electrons. Other sources as electrons from $\pi^0$- and $\eta$-Dalitz
decays, photo-conversions, and decays of light vector-mesons ($\rho$,
$\phi$) turned out to be neglectable, since \textit{(i)} their average
\pt\ is too low and \textit{(ii)} their rate is small compared to the
abundance of misidentified hadrons. In addition, most electrons from
photo-conversion can easily be rejected by requiring that all tracks
point back to the primary event vertex.

The most crucial factor in the overall background rejection is the
hadron rejection power (e/h) of the EMC.  However, for $p <
2\ \gevc$, a region where the hadron rejection capabilities of the EMC
are degraded, the combined dE/dx information of SVT and TPC helps to
augment the S/B ratio considerably.  It is important to note, that the
hadron rejection enters quadratically into the background. 

The magnitude of the background depends strongly on the momentum cut
applied to the electron candidates. In order to access the low-\pt\ 
\jpsi\ region STAR has chosen m$_{\mathrm{J}/\psi}/2 \simeq 1.5\ \gevc$
as the lowest value to study.  The higher the cut the better the e/h
rejection from the EMC and the lower the charged pion yield.

\begin{table}[hbt]
\begin{tabular}{|l|c|c|c|c|} 
\hline
cut &  \jpsi-yield   & S/B & $\sigma$ after $10^7$ sec &
$\sigma$ after $10^6$ sec \\ \hline
p$_e > 1.5\ \gevc$  &   40 k         & 1:3 &  76 & 24  \\ \hline    
p$_e > 2.0\ \gevc$  &  10 k          & 3:1 &  77 & 24  \\ \hline    
\end{tabular}
\label{tab:rates}
\caption{Current estimates on \jpsi\ yields, signal-to-background ratio,
    and statistical significance of the signal in the STAR experiment.}
\end{table}

Before any measurement can be addressed it is important to estimate
the achievable yields and the statistical significance of the signal
under realistic assumptions.  Table \ref{tab:rates} summarizes the
resulting yields, the signal-to-background ratio, and the statistical
significance after $10^7$ and $10^6$ sec. Note, that a nominal RHIC
run has $10^7$sec.  Here one assumes 100\% level-3 trigger efficiency
running at the design rate of 100 Hz, and full EMC coverage including
pre-shower detectors. The estimate includes the reconstruction
efficiencies of the various detectors. Known cross-section from
elementary collisions were used to exptrapolate to Au+Au at \sqrtsNN\,
= 200 GeV.  Two different sets of cuts on the electron momenta are
shown in order to demonstrate its effect on the S/B ratio.
Interestingly, the higher S/B ratio for the larger \pt\ cut balances
the decreased signal strength, thus resulting in the same statistical
significance for both cases.

For the actual measurement one has to study \jpsi\ production not only
in the most central collisions, but has to vary the centrality and
possibly the colliding systems. Although the \jpsi\ yield will
decrease for semi-central and peripheral collisions the
signal-to-background ratio will decrease even stronger (almost
quadratically) which will ease the extraction of the signal
significantly.

\section{Summary and Conclusions}
The workshop was a wonderful forum for learning and exchanging new
ideas about the physics relevant at RHIC.  There is lots of exciting
physics in a field with much potential for discovery.  We wish to
thank the organizers and all the students for their active
participation and thought provoking questions.


\theendnotes

\printindex
\label{lastpage}
\end{article}

\addcontentsline{toc}{section}{References}

\begin{thebibliography}{99}

\bibitem{hagedorn} R.~Hagedorn, Suppl.~A.~Nuovo Cimento VolIII, No.2 (1965) 150. 
\bibitem{pbm-sps} P.~Braun-Munzinger, I.~Heppe, and J.~Stachel,  Phys.~Lett.~B465 (1999) 15.
\bibitem{pbm-rhic} P.~Braun-Munzinger {\it et al.},  Phys.Lett.~B518 (2001) 41-46.
\bibitem{satz}T.~Matsui and H.~Satz, Phys.~Lett.~B178,~416 (1986).
\bibitem{mueller}R.~Baier, Y.L.~Dokshitzer, A.H.~Mueller and D.~Schiff, Phys.~Rev.~C58, 1706 (1998); hep-ph/9803473.
\bibitem{mueller2}R.~Baier, Y.L.~Dokshitzer, A.H.~Mueller, S.~Peigne and D.~Schiff, Nucl.~Phys.~B483, 291 (1997); hep-ph/9607355.
\bibitem {STAR} Conceptual Design Report for the Solenoidal Tracker At RHIC, The STAR Collaboration, PUB-5347 (1992); J.W.~Harris \textit{et al.}, Nucl.~Phys.~A 566, 277c (1994).
\bibitem {PHENIX} PHENIX Experiment at RHIC - Preliminary Conceptual Design Report, PHENIX Collaboration Report (1992).
\bibitem {BRAHMS} Interim Design Report for the BRAHMS Experiment at RHIC, BNL Report (1994).
\bibitem {PHOBOS} RHIC Letter of Intent to Study Very Low pt Phenomena at RHIC, PHOBOS Collaboration (1991).
\bibitem {Spin} Proposal on Spin Physics Using the RHIC Polarized Collider, RHIC Spin Collaboration (1992).
\bibitem{phobos-prl1}B.B.~Back {\it et al.} (PHOBOS), Phys.~Rev.~Lett.~85, 3100 (2000); hep-ex/0007036.
\bibitem{phenix-nch}K.~Adcox {\it et al.} (PHENIX), Phys.~Rev.~Lett.~86, 3500 (2001); nucl-ex/0012008.
\bibitem{brahms-qm}F.~Videbaek {\it et al.} (BRAHMS), Proceedings to the 15th International Conference on Ultrarelativistic Nucleus-Nucleus Collisions 2001. 
\bibitem{star-qm}J.~Harris {\it et al.} (STAR), Proceedings to the 15th International Conference on Ultrarelativistic Nucleus-Nucleus Collisions 2001. 
\bibitem{phobos-nch2}B.B.~Back {\it et al.} (PHOBOS); nucl-ex/0105011.
\bibitem{ekrt}K.J.~Eskola, K.~Kajantie, and K.~Tuominen, Phys.~Lett.~B497, 39 (2001); hep-ph/0009246.
\bibitem{hijing}X.N.~Wang and M.~Gyulassy, Phys.~Rev.~Lett.~86, 3496 (2001); nucl-th/0008014.
\bibitem{bjorken} J.D.~Bjorken, Phys.~Rev.~D, Vol.~27, page 140 (1983).
\bibitem{brahms-ratio}I.G.~Bearden {\it et al.} (BRAHMS); nucl-ex/0106011.
\bibitem{phenix-et}K.~Adcox {\it et al.} (PHENIX), Phys.~Rev.~Lett.~87, 052301 (2001); nucl-ex/0104015.
\bibitem {olliCorsica} See contribution of J.-Y.~Ollitrault in these proceedings.
\bibitem{sorge} H.~Sorge, Phys.~Lett.~{\bf B}402, 251 (1997).
\bibitem{olli92} J.-Y.~Ollitrault, Phys.~Rev.~D {\bf 46}, 229 (1992).
\bibitem{e877flow2} E877 Collaboration, J.~Barrette {\it et al.}, Phys.~Rev.~C {\bf 55}, 1420 (1997).
\bibitem{e895} E895 Collaboration, C.~Pinkenburg {\it et al.}, Phys.~Lett.~{\bf 83}, 1295 (1999).
\bibitem{na49prl} NA49 Collaboration, H.~Appelsh\"{a}user {\it et al.}, Phys.~Lett.~{\bf 80}, 4136 (1998).
\bibitem{na49flow} A.M.~Poskanzer and S.A.~Voloshin for the NA49 Collaboration, Nucl.~Phys.~{\bf A661}, 341c (1999).
\bibitem{wa98} WA98~Collaboration, M.M.~Aggarwal et al., Phys.~Lett.~{\bf B}403, 390 (1997); M.M.~Aggarwal {\it et al.}, Nucl.~Phys.~{\bf A638}, 459 (1998).
\bibitem{starflow} STAR~Collaboration, K.H.~Ackermann {\it et al.}, Phys.~Lett.~{\bf 86}, 402 (2001).
\bibitem{physeflow} S.A.~Voloshin and A.M.~Poskanzer, Phys.~Lett.~{\bf B}474, 27 (2000).
\bibitem{uliflowPRC} P.F.~Kolb, J.~Sollfrank, and U.~Heinz, Phys.~Rev.~C {\bf 62}, 054909 (2000).
\bibitem{hbt} R.~Hanbury Brown and R.Q.~Twiss, Phil.~Mag.~{\bf 45} (1954) 663.
\bibitem{hbtreview} U.A.~Wiedemann and U.~Heinz, Rhys.~Rept.~{\bf 319} (1999) 145.,
\bibitem{makhlin_sinyukov} A.~Makhlin and Y.~Sinyukov, Z.~Phys.~{\bf C39} (1998) 69.
\bibitem{star_hbt_paper} C.~Adler \textit{et al.}, Phys.~Rev.~Lett.~87, 082301 (2001).
\bibitem{Pratt} S.~Pratt, Phys.~Rev.~D {\bf 33}, 1314 (1986); G.~Bertsch, M.~Gong and M.~Tohyama, Phys.~Rev.~C{\bf 37}, 1896 (1988); and G.~Bertsch, Nucl.~Phys.~A{\bf498}, 151c (1989).
\bibitem{Rischke} D.~H.~Rischke, Nucl.~Phys.~A610 (1996) 88c; D.H.~Rischke and M.~Gyulassy, Nucl.~Phys.~A608 (1996) 479.
\bibitem{johnson} S.C.~Johnson, nucl-ex/0104020.
\bibitem{soff_bass_dumitru} S.~Soff, S.A.~Bass and A.~Dumitru, nucl-th/0012085.
\bibitem{phx-qm}W.A.~Zajc {\it et al.} (PHENIX), Proceedings to the 15th International Conference on Ultrarelativistic Nucleus-Nucleus Collisions 2001; nucl-ex/0106001.
\bibitem{wang-prc}X.N.~Wang, Phys.~Rev.~C61, 064910 (2000); nucl-th/9812021.
\bibitem{star-highpt}C.~Adler {\it et al.} (STAR); nucl-ex/0106004.
\bibitem{drees} P.~Braun-Munzinger \textit{et al.}, Eur.~Phys.~J.~C1 (1998) 123-130; nucl-ex/9704011.
\end{thebibliography}
\end{document}